\pdfoutput=1

\documentclass[10pt]{article}
\scrollmode
\usepackage{amsmath,amssymb}
\usepackage{eucal}
\usepackage[dviwindo]{graphicx}
\usepackage[dviwindo]{graphics}

\numberwithin{equation}{section}
\newcommand{\ben}{\begin{eqnarray}}
\newcommand{\een}{\end{eqnarray}}
\newcommand{\la}{\label}

\begin{document}

\title{Classes of Exact Solutions to Regge-Wheeler and Teukolsky  Equations}

\vskip 1.5truecm

\author{P.~P.~Fiziev\thanks{Department of Theoretical Physics, University of Sofia,
Boulevard 5 James Bourchier, Sofia 1164, Bulgaria,
E-mail:\,\,\,fiziev@phys.uni-sofia.bg}}

\date{}
\maketitle

\begin{abstract}

The Regge-Wheeler equation describes axial perturbations of
Schwarzschild metric in linear approximation.
Teukolsky Master Equation describes perturbations of Kerr metric in the same approximation.
We present here  unified description of all classes of exact solutions to these equations
in terms of the confluent Heun's functions.
Special attention is paid to the polynomial solutions,
which yield novel applications of Teukolsky Master Equation
for description of relativistic jets and astrophysical explosions.
\end{abstract}

%
\sloppy
\scrollmode

\section{Introduction}
At present the study of different type of perturbations of the gravitational field of
black holes (BH), neutron stars (NS) and other compact astrophysical objects
is a very active field for analytical, numerical, experimental and
astrophysical research. Ongoing and nearest future experiments,
based on perturbative and/or numerical analysis of relativistic
gravitational dynamics, are expected to provide critical tests of the
existing theories of gravity \cite{KipTorn}.

In the last five years
the sensitivity of the operating detectors for gravitational waves LIGO, VIRGO, GEO, TAMMA
has been improving at a formidable rate and one may expect the
first direct observation of gravitational waves in the nearest future.
Note that the existing observational results, collected in the last five years,
give only limitations on the number of the BH-BH, BH-NS and NS-NS
mergers. Since such mergers are still not observed, this number seems to be below
the optimistic theoretical expectations, announced some eight years ago\footnote{Up
to two detections of BH-BH mergers per year for LIGO I have been forecasted at that time.}.
The number of real mergers is believed to be consistent with the recent
theoretical and observational constraints \cite{BBH}.
The already started large projects like advanced LIGO and especially LISA, hopefully will
bring into being the gravitational wave astronomy in the next
decade. Thus we are expecting to discover new fundamental physics.

Another outstanding physical problem is presented by the gamma ray bursts (GRB) -- the
most powerful explosions in our universe after the Big Bang and the
relativistic jets, related with them, as well as with other astrophysical objects.
Due to recent developments
of gamma ray astronomy in space missions SWIFT, Chandra, Huble Space
Telescope, Spitzer, HETTE-2, BeppoSAX, and AGILE, together with ground observations by ESA
and many other observatories, we already have very good observational data, which is
still waiting for adequate theoretical explanation. The recently started
Fermi/GLAST mission will give us more complete and precise data in the nearest
future. Concerning the theoretical situation one has to stress that the existing theoretical models of GRB
do not give a clear and acceptable explanation of the observational facts \cite{GRB}.
Moreover, there is a kind of crisis in this area, since the observational data seem to contradict the
old models of central engine of long GRB. In addition, the presence of BH in short GRB was recently refuted
by the existing detectors of gravitational waves \cite{Andromeda}.

New physical effects, due to the rotation of the gravitational field described by general relativity (GR),
may play role in the supernova explosions. Mathematical tools for the study of the generation
of waves with different spins during supernova explosions are needed, too.

A basic approach for theoretical study of the above phenomena is the perturbation theory
of different background gravitational fields. At present we have two versions of this theory. One is based
on direct perturbations of the space-time metric. The other one is based on perturbations of the
scalars of Riemann curvature tensor, constructed using a specific tetrad in the space-time.
In the present article we are considering the exact solutions of the perturbation equations in both versions.
One of our goas is to compare the two sets of basic results.

The well known Regge-Wheeler Equation (RWE) \cite{RW}
\ben  \partial_t^2\, {}_s\Phi_{l} +\left(-\partial^2_{x}
+{}_sU_l\right){}_s\Phi_{l} = 0 \la{RW}\een
describes in linear approximation the axial perturbations of Einstein equations, using
the Schwarzschild metric as a background. It plays an important role in modern perturbation
treatment of Schwarzschild black hole (SBH) physics. Its study has a
long history and significant achievements \cite{Chandra, QNM}.

Usually one works only with bounded solutions of the corresponding angular equation \cite{RW,QNM}.
Then the effective potential in the radial RWE (\ref{RW})
${}_sU_{l}(r)=\left(1-{{1}\over r}\right)\left({{l(l+1)}\over {r^2}}
+{{1-s^2}\over {r^3}} \right)$,
has a specific dependence: $l(l+1)$ on an integer $l\geq |s|$.
The area radius $r\geq 0$ can be expressed explicitly as a function
of the "tortoise" coordinate
$x\!=\!r+r_{\!{}_{Sch}}\ln\left(|r/r_{\!{}_{Sch}}-1|\right)$
using the Lambert-W function: $r=\text{LambertW}(\pm e^{x-1})+1$. In the last formula
the sign $"+"$ stands for the SBH exterior $r\in (1,\infty)$ and the sign $"-"$
-- for the SBH interior $r\in (0,1)$. Hereafter we are using units
in which the Schwarzschild radius $r_{\!{}_{Sch}}\!=\!2M\!=\!1$. In the potential ${}_sU_{l}(r)$
the quantity $s$ has values $s=0, \pm1, \pm2$.
The most important from astrophysical point of view are the cases $s\!=\pm1$ and $s\!=\pm2$,
which describe electromagnetic and gravitational waves, correspondingly.

The study of perturbations of the Schwarzschild metric was started
in 1957 by Regge and Wheeler and was developed essentially by
Chandrasekhar, Leaver and many others, see in \cite{RW, Chandra, Leaver}, in the review
articles \cite{QNM} and in the large amount of references therein. Especially,
analytical study of the solutions was started in \cite{Leaver}
and extended by different approximate methods, see in \cite{QNM}. The
exact analytical solutions of this problem have been found recently \cite{F}.
Using them one obtains the quasi-normal modes (QNM) of Schwarzschild metric
under {\em different} boundary conditions in the most natural and
straightforward way -- solving numerically the corresponding
boundary problems formulated in terms of the exact solutions.
These solutions are useful both for a more deep understanding of the
corresponding physical problems and for formulation of new ones.
In the present article we give a more general consideration of the Regge-Wheeler
perturbation theory, considering {\em all} types of exact solutions of RWE.
In particular, we pay special attention to the different types of polynomial solutions
to RWE and recover new classes of such solutions.
For this purpose we are to modify and extend to all cases the notions and notations,
used in \cite{F} for detailed description of the solutions.

The ansatz ${}_s\Phi_{l}(t,r)=e^{-i\omega t} {}_sR_{\omega,l}(r) $
with complex frequency $\omega=\omega_R+i\omega_I$ brings
us to the stationary problem in the outer domain $r>1$:
\ben
\partial^2_x\, {}_sR_{\omega,l}+\left(\omega^2-{}_sU_{l}\right)
{}_sR_{\omega,l}=0. \la{R}\een
Its {\em exact} solutions were described in \cite{F} in terms
of the confluent Heun's functions $\text{HeunC}\!\left(\alpha, \beta,
\gamma, \delta,\eta,z \right)$ \cite{Heun}.

In the exterior domain the variable $r$ plays the role of 3D-space
coordinate and the variable $t$ -- the role of the exterior time of a
distant observer. Because of the change of signs of metric's
eigenvalues $\lambda_t=g_{tt}=1-1/r$ and $\lambda_r=g_{rr}=-1/(1-1/r)$, in the interior
domain the former Schwarzschild time variable $t$ plays the role of a
radial space variable $r_{in}=t\in (-\infty,\infty)$ and the area radius
$r\in(1,0)$, plays the role of a time variable. As a result, the
Regge-Wheeler "tortoise" coordinate $x\in (-\infty,0)$ presents a
specific time coordinate in the inner domain. Here for the study of the
solutions of \eqref{RW} it is useful to stretch the interior
time interval to the standard one by further change of the
variable: $x\to t_{in}\!=\!x\!-\!1/x\in (-\infty,\infty)$. Thus we are
placing the already existing singularities at natural places in the corresponding complex plane
and facilitate significantly the numerical calculations \cite{F}.

These comments are important for the physical interpretation of the
mathematical results. In particular, the natural form of the
interior solutions of \eqref{RW} is:
${}_s\Phi_{\omega,l}^{in}(r_{in},t_{in})=e^{-i\omega r_{in}}
{}_sR_{\omega,l}\left(r(t_{in})\right)$, where
$r(t_{in})\!=\text{LambertW}\left(-exp\left( {{t_{in}}\over{2}}\!-\!1\!-\!
\sqrt{\left({{t_{in}}\over{2}}\right)^2\!+\!1}\,\right)
\right)\!\!+\!1$.
The dependence of this solution on the interior radial variable
$r_{in}$ is simple. Its dependence on the interior time $t_{in}$ is
governed by  equation \eqref{R} with interior-time dependent potential
${}_sU_{l}$. Despite of this unusual feature of the solutions in
the SBH interior, using this approach we obtain a basis of functions, which are
suitable for the study of the corresponding linear perturbations.

The negativity of the imaginary part $\omega_I=\Im(\omega)<0$ of the frequency
$\omega$ ensures linear stability of the solutions
with respect to the future time direction $t\to +\infty$ in the exterior domain \cite{RW, NM}.
This condition is not enough to guaranty the stability of the interior domain
(See detailed analysis in terms of the exact solutions to RWE in \cite{F}.).

Usually one is not considering the domain $r<0$,
despite of the fact that it is not excluded by the geometrical meaning of
the area radius $r$. In this domain the variables $t$ and $r$ restore their original meaning.
We have to point out that for the study of the analytical properties of the solutions of RWE
one must consider complex values of the area radius $r\in \mathbb{C}_r$ \cite{F}.

The study of perturbations of rotating relativistic objects in Einstein GR was
pioneered by Teukolsky \cite{Teukolsky} making use of the famous
Teukolsky Master Equation (TME). It describes the perturbations
${}_s\psi(t,r,\theta,\varphi)$ of all physically interesting spin-weights $s=0,
\pm 1/2, \pm 1, \pm 3/2, \pm 2$
to the Kerr background metric in terms of Newman-Penrose scalars.
The pairs of spin-weights $s$ with opposite signs $\sigma=\text{sign}(s)=\pm 1$
correspond to two different perturbations with opposite helicity and spin  $|s|=0, 1/2, 1, 3/2$,
or $2$.
Under proper boundary conditions
for TME one obtains QNM for Kerr black holes (KBH). The various significant results and
references may be found in \cite{Chandra, QNM}.

We have to stress one general feature of the description of rotating relativistic objects.
The Kerr metric describes {\em exactly} the vacuum solution of Einstein equations -- KBH.
The gravitational field outside the rotating compact {\em matter} objects differs form the one,
described by Kerr solution in higher order multipole momenta,
due to the corresponding specific matter distribution (mass distribution and current distribution) \cite{Momenta}.
In the outer domain the contribution of the $l$-th multipole moment to the gravitational field of given object
is of order ${\frac {M}{r}}\!\left({\frac {a}{r}}\right)^l$, $l=0,1,2,\dots$.
Therefore in the outer domain one can fit the gravitational field of {\em any} rotating compact object
by Kerr metric with proper parameters $M$ and $a$. This fitting is {\em exact} for the first two multipole terms.
The difference between the real metric and the Kerr one will appear only in the next terms,
which decrease with the distance as $r^{-l-1}$, $l\geq 2$.
Hence, the effects of the real matter distribution inside the compact
object are negligible at distances greater than several event-horizon-radii and
one can use the Kerr metric outside the event horizon as a very {\em good approximation}
to the gravitational field of rotating compact objects of different nature.
This is the physical basis for our applications of solutions to the Teukolsky Master Equation for
study of different real relativistic objects like massive compact objects, jets, supernova explosions, etc.

On the other hand, the above argument shows that it is impossible to make
a decision about the true nature of the observed massive dark compact objects like ABH,
studying their non-perturbed gravitational field outside the event horizon
and extracting only the parameters $M$ and $a$ from the observational data for the first
two multipole momenta. The same values of the parameters $M$ and $a$
can be prescribed to rotating relativistic objects of different kinds.

The contemporary observational data is not enough
to decide among the possible alternative interpretations
of massive dark compact objects in astrophysics
(See for example the very recent review by Matt Visser in \cite{Visser}.).
We need reliable methods, which are able to make difference between models, like BH, naked singularities,
superspinars, gravastars, boson stars, soliton stars, quark stars, fuzz-balls, dark stars, etc, with the same mass .
These methods have to give indisputable evidences for the real nature of the observed compact dark objects,
currently named astrophysical black holes (ABH).

Such methods, based on the study of spectra of perturbations
of the gravitational field of the real astrophysical objects were proposed in \cite{F, Visser}.
Using them one can "see" directly the event horizon, if it exists,
due to the unique boundary conditions on it.
In \cite{Visser} only approximate methods for study of these problems were discussed.
The exact solutions to the RWE and TME, studied in the present article can help to extend
the approach of \cite{F} to the problems with rotating relativistic objects.

The key feature of the TME is that in Boyer-Lindquist coordinates
one can separate the variables using the ansatz
$\psi(t,r,\theta,\varphi)= e^{-i\omega t} e^{im \varphi}S(\theta)R(r)$,
i.e. looking for solutions in a specific factorized form.
From mathematical point of view the function ${\mathcal K}(t,r,\theta,\varphi)\sim e^{-i\omega t} e^{im \varphi} S(\theta) R(r)$
actually defines a factorized {\em kernel} of integral representation  for the solutions to TME:
\begin{equation}\la{IntRepresentation}
{}_s\psi(t,r,\theta,\phi)=\sum_{m=-\infty}^{\infty}  {1
\over 2\pi} \int\!\!d\omega\int\!\!dE\,
{}_sA_{\omega,E,m}\,e^{-i\omega t}\, e^{im\varphi}\,{}_sS_{\omega,E,m}(\theta)\,{}_sR_{\omega,E,m}(r),
\end{equation}
basically introduced in the problem at hand for the first time in \cite{Teukolsky}.
The study of the QNM \cite{QNM} showed that this kernel can
be singular with respect to the variables $E$ and $r$: it is proportional to Dirac $\delta (E-E_{sp})$, where $E_{sp}$
belongs to some specific for the given problem spectrum; with respect to variable $r$
the kernel may be singular at infinity and at the horizons.
In the existing literature only regular with respect to the variable $\theta$ kernels are in use.
In the present article we are starting the consideration both of regular and singular with respect
to $\theta$ kernels in the natural integral representation \eqref{IntRepresentation} of solutions to the TME.
The different type of kernels are to be used for solution of different boundary problems.
Note that from physical point of view is important the regularity of the very solution
${}_s\psi(t,r,\theta,\phi)$ in equation \eqref{IntRepresentation}.
The kernels like ${\mathcal K}(t,r,\theta,\varphi)$
are an auxiliary mathematical objects. One is often forced to use singular kernels
for the natural integral representations of the solution of physical problems.
The regularity of the function ${}_s\psi(t,r,\theta,\phi)$
with respect to the variable $\theta$ depends on the choice of the amplitudes ${}_sA_{\omega,E,m}$.

As a result of factorization of the kernel ${\mathcal K}(t,r,\theta,\varphi)$  a pair of two connected
ordinary differential equations for its nontrivial factors ${}_sS_{\omega,E,m}(\theta)$
and ${}_sR_{\omega,E,m}(r)$ arises -- the Teukolsky
angular equation (TAE) \cite{Teukolsky, angular}
\begin{subequations}\la{angE:ab}
\ben
{\frac {1}{\sin \theta}} {\frac {d}{d\theta}} \left( \sin
\theta {\frac {d} {d\theta}}\,{}_sS_{\omega,E,m}( \theta ) \right)
+{}_sW_{\omega,E,m}(\theta) {}_sS_{\omega,E,m}(\theta)\!=0, \hskip 2.2truecm \la{angE:a}\\
{}_sW_{\omega,E,m}(\theta)=E+a^2\omega^2\cos^{2}\theta-2sa\omega \cos
\theta -(m^2+s^2+2ms\cos\theta)/\sin^2 \theta; \la{angE:b}
\een
\end{subequations}
and the Teukolsky radial equation (TRE) \cite{Teukolsky}

\begin{subequations}\la{radE:ab}
\ben
{\Delta}^{-s}{\frac {d }{dr}} \left( {\Delta}^{s+1}
{\frac{d}{dr}}\,{}_sR_{\omega,E,m}(r)  \right) + {}_sV_{\omega,E,m}(r)\, {}_sR_{\omega,E,m}(r) =0, \hskip 0truecm
\la{radE:a}\\
{}_sV_{\omega,E,m}(r)={\frac 1 \Delta}{{K}^{2}} -is {\frac 1 \Delta} {\frac {d\Delta} {dr}}K-L.\hskip 2truecm
\la{radE:b}
\een
\end{subequations}
Here the azimuthal number $m$ has arbitrary integer values $m=0, \pm 1, \pm 2,\dots $, and
$\Delta=r^2-2Mr+a^2$, $K=\omega(r^2+a^2)-ma$, $L=E - s(s+1) + a^2\omega^2 -2ma\omega - 4\,is\omega\,r$.
The standard real parameter $a$ is related with the angular momentum of the Kerr metric,
$M$ is the Keplerian mass of the Kerr solution.
The two {\em complex} parameters $\omega$ and $E$ -- the constants of the separation,
are to be determined using the boundary conditions of the problem.

As in the case of Schwarzschild background, the negativity of the imaginary part
$\omega_I=\Im(\omega)<0$ of the frequency $\omega=\omega_R+i\omega_I$ may
ensure linear stability of the solutions in the exterior domain of the Kerr metric
with respect to the future time direction $t\to +\infty$ \cite{Teukolsky, Kerr_Stability}.
In the interior domain the solutions to TME are not stable \cite{Kerr_Instability}.

Much like in case of the Schwarzshild solution,
in the interior of the Kerr metric -- between the zeros $r_{\pm}$: $0\leq r_{-}< r_{+}$
of the function $\Delta$, two of the eigenvalues: $\lambda_{t}$ and $\lambda_{r}$ of the metric in Boyer-Lindquist
coordinates simultaneously change their signs. Indeed, one pair of eigenvalues are
$\lambda_{\theta}=g_{\theta\theta}=r^2+a^2\cos^2\theta$ and $\lambda_{r}=g_{rr}=(r^2+a^2\cos^2\theta)/\Delta$.
The second pair of eigenvalues are the roots $\lambda_{t}, \lambda_{\phi}$ of the equation
$\lambda^2-(g_{tt}+g_{\phi\phi})\lambda+g_{tt}g_{\phi\phi}-g_{t\phi}^2$.
Their product equals $\lambda_{t}\lambda_{\phi}=-\Delta\sin^2\theta$.
The last expression, together with the form of  $g_{rr}$ proves the simultaneous change of the signs of the
two eigenvalues $\lambda_{t}, \lambda_{r}$, when the variable $r$ crosses the horizons $r_{\pm}$,
since the determinant of the metric  $g=-(r^2+a^2\cos^2\theta)^2\sin^2\theta$ does not vanish there.
As a result, between the two horizons $r_{\pm}$ the variable $t_{in}=x\in (-\infty,\infty)$
plays the role of the interior time and the variable $r_{in}=t$ is the interior radial variable.
We are using the following  Kerr-metric-tortoise-coordinate:
$x=r+a_{+}\ln|r/r_{+}-1|-a_{-}\ln|r/r_{-}-1|\in (-\infty,\infty)$, where $a_\pm={\frac{r_{+}+r_{-}}{r_{+}-r_{-}}}r_{\pm}$.
It is a straightforward generalization of the proposed in \cite{Teukolsky} tortoise variable for the exterior domain $r\in (r_{+},\infty)$.
Since our expression is valid in the interior domains, too, the inverse function defines $r=r(t_{in})$ when $r\in (r_{-}, r_{+})$.
In the second interior domain $r<r_{-}$ the variables $r$ and $t$ restore their original meaning.
For a detailed analysis of the light cones in Kerr geometry see \cite{LC}.

Despite of the essential progress both in the numerical study \cite{KerrQNM_numeric} of the
solutions to the  equations (\ref{angE:a}) and (\ref{radE:a})
and in the investigation of their analytical properties \cite{KerrQNM_analytic},
at present there is a number of basic questions remaining unanswered.
For example, it is well known for a long time \cite{TME_Heun} that RWE (\ref{R}), as well as
TAE (\ref{angE:a}) and TRE (\ref{radE:a}) can be reduced to the confluent Heun
differential equation \cite{Heun}
\begin{align}
H''+\left(\alpha+{{\beta+1}\over{z}}+{{\gamma+1}\over{z-1}}\right)H'+
\left( {\mu\over z}+{\nu\over{z-1}} \right)H = 0.\la{DHeunC}
\end{align}
Here the constants $\mu$ and $\nu$ are related with the accepted in the notation
$\text{HeunC}(\alpha,\beta,\gamma,\delta,\eta,z)$
ones: $\alpha, \beta, \gamma, \delta, \eta $ according to the
equations
\begin{subequations}\la{munu:ab}
\ben
\delta =\mu+\nu - \alpha{\frac{\beta+\gamma+2} 2},\la{munu:a}\\
{\eta={\frac{\alpha(\beta+1)} 2}-\mu-\frac{\beta +\gamma+\beta\gamma} 2}.\la{munu:b}
\een
\end{subequations}
To the best of our knowledge the exact analytical solutions of the
angular and radial  equations \eqref{angE:a} and \eqref{radE:a} are still not
described in the literature in terms of confluent Heun function
$\text{HeunC}(\alpha,\beta,\gamma,\delta,\eta,z)$ -- the unique particular
solution of the \eqref{DHeunC}, which is regular in the vicinity of
the regular singular point $z=0$ and normalized by the requirement
$\text{HeunC}(\alpha,\beta,\gamma,\delta,\eta,0)=1$
(See for details \cite{Heun}.)\footnote{
In the present article we are using the Maple-computer-package notations for Heun functions.
Basically, these notations are borrowed from the two mile-stone articles
on modern theory of Heun functions by Decarreau et al. in \cite{Heun}, and at present seem to be
most popular, since the Maple package is the only one for analytical and numerical
work with Heun functions.}.

In the late 2006 a program for filling this gap has been started
as a natural extension of the articles \cite{F}. The first results were quite stimulating \cite{PFDS},
but serious difficulties came across both in analytical and in numerical studies.
This is because the theory of Heun's functions,
as well as numerical tools for computer calculations with them are still not developed enough.
The main purpose of the present series of articles is to report some of the basic results, obtained for detailed
description of the exact solutions and to describe different boundary
problems for RWE, TRE, TAE, and TME in terms of the confluent Heun function,
developing both the theory and computational tools for this function.

We outline several applications of these solutions to well known physical problems like QNM of static and rotating BH,
as well as applications to other astrophysical phenomena.
In particular, we found novel simple mathematical description of relativistic jets, applying specific solutions of TME,
which were not used up to now \cite{PFDS_astroph:HE}. We have to stress that the relativistic jets are
quite common phenomenon at very different physical scales \cite{Jets}.
Around very different astrophysical objects: small mass brown dwarfs,
neutron stars, protostars, GRB, quazars, blazars, galactic centers, and even around galactic clusters
have been observed jets of various size, corelated with the scales of these objects.
Therefore, if possible, a common description and models of relativistic jets are highly desirable.

According to \cite{Heun} the confluent Heun function $\text{HeunC}(\alpha,\beta,\gamma,\delta,\eta,z)$
reduces to a polynomial of degree $N\geq 0$ of the variable $z$,
if and only if the following two conditions are satisfied:
\begin{subequations}
\la{PolynomCond:ab}
\ben
{\frac{\delta}{\alpha}}+{\frac{\beta+\gamma}{2}}+N+1=0,\la{PolynomCond:a}\\
\Delta_{N+1}(\mu)=0.\la{PolynomCond:b}
\een
\end{subequations}
Further on we call the first condition, i.e., the \eqref{PolynomCond:a}
-- a "$\delta$-condition", and the second one, i.e., the \eqref{PolynomCond:b} -- a "$\Delta_{N+1}$-condition".

We represent the three-diagonal determinant $\Delta_{N+1}(\mu)$ in the following specific explicit form:
\ben
\!\left|\!
\begin{array}{ccccccc}
\mu\!-\!q_{1}\! & 1(1\!+\!\beta) & 0 &\!\ldots\! & 0 & 0 & 0\\
N\alpha & \mu \!-\!q_{2}\!+\!1\alpha & 2(2\!+\!\beta) & \!\ldots\! & 0 & 0 & 0  \\
0 & (N\!-\!1)\alpha & \mu\!-\!q_{3}\!+\!2\alpha & \!\ldots\!& 0 & 0 & 0 \\
\vdots & \vdots & \!\vdots\! &\!\ddots\! &\vdots &\vdots &\vdots \\
0 & 0 & 0 & \!\ldots\! & \mu \!-\!q_{N\!-\!1}+(N\!-\!2)\alpha & (N\!-\!1)(N\!-\!1\!+\!\beta) & 0 \\
0 & 0 & 0 & \!\ldots\! & 2\alpha & \mu \!-\!q_{N}\!+\!(N\!-\!1)\alpha & N(N\!+\!\beta)\\
0 & 0 & 0 & \!\ldots\! & 0 & 1\alpha & \mu \!-\!q_{N\!+\!1}\!+\!N\alpha
\end{array}
\!\right|,
\la{Delta}
\een
which turns to be useful for calculations. Here $q_{n}=(n-1)(n+\beta+\gamma)$.

On the other hand, the so called algebraically special solutions of RWE and TRE
were discovered long time ago \cite{AlgSpetial}.
These are of a generalized polynomial type. According to the existing literature
these solutions describe pure incoming or pure outgoing waves.
To the best of our knowledge the algebraically special solutions
are still not discussed in terms of Heun polynomials.

As far as we know, applications of this class of solutions to real physical problems
can not be found in the existing literature on gravitational physics.
If such waves are spreading only in one direction, they seem to be most suitable
for description of relativistic jets and supernova explosions.

We have to stress one more argument in this direction.
Very recently the algebraically special solutions of RWE and TME were proved to be
relevant for the study of instabilities of different kind of some more or less "exotic"
solutions to the Einstein equations \cite{ASM_non_stabilities}.
Obviously, the physical manifestation of the instabilities of the mathematical solutions
are the explosions of the corresponding objects. Therefore it seems natural to look for a description
of astrophysical explosions in terms of polynomial solutions of TME, which are stable in the future and
instable in the past.

Our study of the polynomial solutions of TAE, which were completely ignored in the gravitational physics
up to now, shows that these indeed may describe
in a most natural way the collimation of the jets and demonstrate reach variety of jet forms,
very similar to the observed ones \cite{PFDS, PFDS_astroph:HE}.
We are going to analyze these problems on a correct mathematical basis using the confluent Heun functions.

It is well known that the Kerr metric approaches the non-rotating Schwarzschild one in the limit  $a \to 0$.
One of our aims is to stress the analogy and the differences between the solutions of RWE and TME.
These two equations describe the perturbations of Schwarzschild and Kerr metrics
in terms of {\em different} quantities: RWE -- in terms of direct perturbations of metric
and TME -- in terms of perturbation of Weyl scalars.
Therefore in the limit $a \to 0$ there exist a smooth transition  from perturbations
of KBH to perturbations SBH in terms of Weyl scalars,
but a simple transition from the solutions of TME to the solutions of RWE is not possible \cite{Chandra}.
Nevertheless, the mathematical analogy between the corresponding solutions becomes quite transparent
when the solutions are represented in terms of confluent Heun functions.

Another point of ours is to collect at one place the known facts, together with the new results in the field
and to describe all of them uniformly and in common terms and notations. This way we hope to obtain a more clear picture
of the quite complicated present-day state of the arts in the perturbation theory under consideration and its possible further
developments.

The large amount of the accomplished work and the number of the obtained results forces us to publish them
in a series of articles, starting with this one.

\section{Exact Solutions to the Regge-Wheeler Equation in Terms of the Confluent Heun's functions}

In area-radius-variable $r$ the RWE (\ref{R}) reads:
\ben \la{P} {{d^2\,{}_sR_{\omega,l}}\over{dr^2}}-
\left({1\over{r}}+ {1\over{1-r}}\right){{d\,{}_sR_{\omega,l}}\over{dr}}+\hskip 8.truecm\\
\Bigg(
\omega^2+\big({l(l+1)-s^2+1}\big)
\left({1\over{r}}+ {1\over{1-r}}\right)-{{s^2-1}\over{r^2}}-{{2\omega^2}\over{1-r}}+{{\omega^2}\over{(1-r)^2}}
\Bigg){}_sR_{\omega,l}=0.\nonumber
 \een

The anzatz ${}_sR_{\omega,l}(r)=r^{s+1}(r-1)^{i\omega} e^{i\omega
r}H(r)$ reduces it to the confluent Heun equation \eqref{DHeunC} with the following specific parameters:
$\alpha=2i\omega, \beta=2s, \gamma=2i\omega, \delta=2\omega^2, \eta=s^2-l(l+1)$ \cite{F}.
This anzatz was used for the first time for analytical and numerical study of the solutions
of RWE by Leaver in \cite{Leaver} without reference to Heun equation and Heun functions.

The equation (\ref{R}) has three
singular points in the whole complex plane $\mathbb{C}_r$ \cite{F, Heun}.
Two of them: $r=0$ and $r=1$ are regular and can be treated on equal
footing. The third one: $r=\infty$ is an irregular singular point.

Note that, after all, the horizon $r=1(=2M)$ turns out to be a singular point
for the perturbations of SBH, despite of the fact that it is not a singularity of the algebraic
invariants of the Riemann's curvature tensor ${\cal R}_{ijkl}$.
The algebraic invariants are able to indicate only the curvature singularity at the point $r=0$.
The horizon does not define a singular surface of these invariants in the Schwarzschild space-time manifold
$\mathbb{M}^{(1,3)}$ and usually is treated as a pure coordinate singularity
of the metric in Schwarzshild coordinates.
Since the {\em algebraic} invariants
do not fix the whole geometry of the space-time manifold, their consideration is not enough to
recover all space-time properties.
For this purpose one must consider a large enough number of high-order-differential-invariants
of the Riemann's tensor ${\cal R}_{ijkl }$ \cite{Cartan}. For example, the differential invariant of first order
\ben
DI=-\left(\nabla \ln r\right)^2=
-{\frac{1}{36}}\left(\nabla \ln\left({\cal R}_{ijkl}{\cal R}^{ijkl}\right)\right)^2=
{\frac{1}{r^2}}\left(1-{\frac{2M} r}\right)
\la{diffInvSBH}\een
can be used as a coordinate-independent indicator both of the event horizon $r=2M (=1)$ of SBH
and of the singularity at $r=0$. Obviously, both of the last two values of the area variable
$r$ define non-coordinate geometrical objects in the Schwarzschild space-time manifold $\mathbb{M}^{(1,3)}$
and at the same time -- alike singularities of equation \eqref{P}.
It becomes clear that for the study of small deviations from the background metric it is necessary
to consider differential invariants of the Riemann's tensor,
relevant to corresponding variations of geometry.

Using the confluent Heun function one can write down 16 local Frobenius type solutions of the RWE (\ref{P}):
\ben
{}_sR^\pm_{\omega,l,\sigma_\alpha,\sigma_\beta,\sigma_\gamma}(r)/r=e^{\sigma_\alpha{\frac{\alpha_{{}_\pm} z_{{}_\pm}} 2}} z_{{}_\pm}^{\sigma_\beta{\frac{\beta_{{}_\pm}} 2}} z_{{}_\mp}^{\sigma_\gamma{\frac{\gamma_{{}_\pm}} 2}}
\text{HeunC}({\sigma_\alpha\alpha_{{}_\pm},\sigma_\beta\beta_{{}_\pm},\sigma_\gamma\gamma_{{}_\pm},\delta_{{}_\pm},\eta_{{}_\pm},z_{{}_\pm}}).
\la{RWE16local}
\een
Here
\begin{subequations}\ben
\label{RWparameters:abcd}
\alpha_{{}_{+}}\!=2i\omega,\,\, \beta_{{}_{+}}=2s,
\gamma_{{}_{+}}=2i\omega, \delta_{{}_{+}}=2\omega^2,\,\, \eta_{{}_{+}}=\,\,s^2-l(l+1); \la{RWparameters:a}\\
\alpha_{{}_{-}}\!\!=\!-2i\omega, \beta_{{}_{-}}\!\!=\!2i\omega,
\gamma_{{}_{-}}\!\!=\!2s, \delta_{{}_{-}}\!\!=\!-2\omega^2, \eta_{{}_{-}}\!\!=\!\!2\omega^2\!+\!s^2\!-\!l(l\!+\!1); \la{RWparameters:b}\\
\sigma_\alpha=\pm 1,\sigma_\beta=\pm 1,\sigma_\gamma=\pm 1\la{RWparameters:c};\\
z_{+}=r,\,\,\,z_{-}=1-r.\la{RWparameters:d}
\een
\end{subequations}

According to equations \eqref{RWE16local} and \eqref{RWparameters:d} the behavior of the solutions
${}_sR^{\pm}_{\omega,l,\sigma_\alpha,\sigma_\beta,\sigma_\gamma}(r)$
around the corresponding singular points $z_{\pm}=0$ is defined by the dominant factor
$z_{{}_\pm}^{\sigma_\beta \beta_{\pm}/2}$.
All other factors in \eqref{RWE16local} are regular around these points.
In contrast, the solutions ${}_sR^{\pm}_{\omega,l,\sigma_\alpha,\sigma_\beta,\sigma_\gamma}(r)$ are in general
singular around the corresponding singular points  $z_{\pm}=1$, i.e., around $r=r_{+}$
-- for ${}_sR^{+}_{\omega,l,\sigma_\alpha,\sigma_\beta,\sigma_\gamma}(r)$
and around $r=r_{-}$ -- for ${}_sR^{-}_{\omega,l,\sigma_\alpha,\sigma_\beta,\sigma_\gamma}(r)$.
This explains the meaning of the upper $\pm$ signs in the notation $R^\pm$.

Only two of the sixteen solutions  (\ref{RWE16local}) are linearly independent.
For different purposes one can use different pairs of independent local solutions.

In addition, using the asymptotic expansion of the confluent Heun function \cite{Heun} we obtain two asymptotic solutions of Tom\`{e} type, i.e.
the local solutions around the irregular singular point $|r|=\infty$ of the RWE in the complex plane
$\mathbb{C}_r$\footnote{The notation $\pm\infty$ in \eqref{RWEinf} indicates
the two different directions on the real $r$-axes in $\mathbb{C}_r$ for approaching the irregular singular point
$|r|=\infty$, i.e. $+\infty$ denotes the limit $r\to +\infty$ and $-\infty$ denotes the limit $r\to -\infty$.
It is consistent with notation $z_{\pm}$, since  $z_{\pm}\to \pm \infty$ when $r\to \infty$.}:
\ben
{}_sR_{\omega,l,\sigma_\alpha,\sigma_\beta,\sigma_\gamma}^{\pm\infty}(r)
\sim{{e^{ i\sigma_{\!\alpha}\,\omega (r + \ln r)}}} \sum_{j\geq 0}{{a_{j}}\left(\pm1\over r\right)^j},\,\,\,\,a_{0}=1.
\la{RWEinf}
\een
For the coefficients $a_{j}=a_{j,\omega,l,\sigma_\alpha,\sigma_\beta,\sigma_\gamma}$  one has a recurrence relation \cite{Heun}
which shows that they increase together with the integer $j$. Hence, the asymptotic series (\ref{RWEinf}) is a divergent one.

\section{Polynomial Solutions to the RWE}
We have to remaind the reader that according to the accepted in \cite{Heun} terminology "polynomial"
are called not only the cases in which the confluent Heun function is indeed a polynomial.
The same terminology is currently in use for all solutions which have a form of finite Taylor series expansion
multiplied by elementary functions.
We shall apply this terminology to the solutions (\ref{RWE16local}) of the RWE and TME, too.
This way we obtain a more reach class of quasi-polynomial solutions \cite{Heun, AlgSpetial}.

The $\delta$-condition (\ref{PolynomCond:a}) gives
\ben
-i\sigma_\alpha\omega^\pm+ \sigma_\beta \left\{{\begin{matrix} s\\i\omega^{-}\end{matrix}}\right\}
 +\sigma_\gamma \left\{{\begin{matrix} i\omega^{+}\\s\end{matrix}}\right\}+N+1=0,\,\,\,N\geq 0;
\la{delta_RWE}
\een
and yields the following two classes of polynomial solutions to the RWE.

\subsubsection{First Class of Polynomial Solutions to the RWE:}

For the solutions ${}_sR^{+}_{\omega,l,\sigma_\alpha,\sigma_\beta,\sigma_\gamma}(r)$
with $\sigma_\alpha=\sigma_\gamma$, and $\sigma_\beta=\pm 1$ equation \eqref{delta_RWE}
gives ${}_sN+1=|s|$, if $\sigma_\beta=-\sigma$ where $\sigma=\text{sign}(s)$. The same result ${}_sN+1=|s|$ we obtain for the solutions
${}_sR^{-}_{\omega,l,\sigma_\alpha,\sigma_\beta,\sigma_\gamma}(r)$ when $\sigma_\alpha=\sigma_\beta$ and
$\sigma_\gamma=-\sigma$. For brevity we will use the short notations:  ${}_sR^{+}_{\omega,l,\sigma_\alpha}(r)={}_sR^{+}_{\omega,l,\sigma_\alpha,-\sigma_s,\sigma_\alpha}(r)$ and
${}_sR^{-}_{\omega,l,\sigma_\alpha}(r)={}_sR^{-}_{\omega,l,\sigma_\alpha,\sigma_\alpha,-\sigma_s}(r)$.

Applying the described in the Introduction general construction, we see that
if in addition the $\Delta_{N+1}$-condition -- \eqref{PolynomCond:b} is fulfilled in the form $\Delta_{|s|}^{\pm}(\mu)=0$,
then the confluent Heun function in \eqref{RWE16local}
reduces to a polynomial of degree $|s|-1\geq 0$.
The $\Delta_{N+1}$-condition gives a finite number $|s|$ different
eigenvalues $\mu_{k=1,\dots,|s|}$ of the parameter $\mu$ for the problem at hand.

1. In the case of RWE in general $|s|=0,1,2$, but the requirement ${}_sN=|s|-1\geq 0$ is satisfied only for $|s|=1,2$.
Hence, first class polynomial solutions of RWE may exist only for electromagnetic ($|s|=1$) and gravitational ($|s|=2$) waves and
there are no such solutions of scalar ($|s|=0$) nature.

2. The explicit form of the  $\Delta_{N+1}$-condition, derived from \eqref{PolynomCond:b} for electromagnetic waves, is
$\Delta^{\pm}_{1}(\mu)=\mu=0$. Then one obtains $l(l+1)=0$, making use of the general relation between $\mu$ and $\eta$
-- \eqref{munu:b}
and the expressions for $\eta_{\pm}$ in \eqref{RWparameters:a} and in \eqref{RWparameters:b}.
Hence, polynomial solutions of the first class for radial RWE with $|s|=1$ and {\em integer}
$l\geq 1$  actually do not exist.
The regularity requirement, usually posed on the solutions of the angular RWE \cite{RW}-\cite{QNM},
is obviously too restrictive for existence of first class electromagnetic waves
with polynomial dependence on the radial variable $r$.

3. The explicit form of the  $\Delta_{N+1}$-condition, derived from \eqref{PolynomCond:b} for gravitational waves, is
\ben
\Delta^{\pm}_{2}(\mu)=\mu^2-2\left(1+{\frac{\sigma_\beta\beta_{\pm}+\sigma_\gamma\gamma_{\pm}}{2}-
{\frac{\sigma_\alpha\alpha_{\pm}}{2}}} \right)\mu-\sigma_\alpha\alpha_{\pm}\left(1+\sigma_\beta \beta_{\pm} \right)=0.
\la{secondRW}
\een
Using the roots of this equations, the \eqref{munu:b},
the expressions for $\eta_{\pm}$ in \eqref{RWparameters:a} and in \eqref{RWparameters:b},
one obtains the following spectrum of the polynomial solutions ${}_sR^{\pm}_{\omega,l,\sigma_\alpha}(r)$:
\ben
{}_{{}_2}\omega^{\pm}_{l,\sigma_\alpha}=\sigma_\alpha{\frac{i}{6}}(l-1)l(l+1)(l+2), \,\,\, l= 2,3,4,\dots
\la{omega_polynomRW}
\een
This spectrum describes stable in the future  ($t\to +\infty$) gravitational waves when $\sigma_\alpha =-1$,
and stable in the past  ($t\to -\infty$) gravitational waves, if $\sigma_\alpha =+1$.

Obviously, using the properties of the confluent Heun function this way we have re-derived the well
known algebraically special solutions to RWE \cite{AlgSpetial},
which describe a specific class of one-way gravitational waves.

\subsubsection{Second Class Polynomial Solutions to RWE:}

Let us impose the $\delta$-condition on the other solutions in \eqref{RWE16local}.
Then because of the asymmetry between the parameters
$\beta$ and $\gamma$ in \eqref{munu:b} we have two different subcases:

A. In the first subcase we consider the solutions ${}_sR^{+}_{\omega,l,\sigma_\alpha,\sigma_\beta,\sigma_\gamma}(r)$
with $\sigma_\alpha=-\sigma_\gamma$. We denote these solutions as  ${}_sR^{+}_{\omega,l,\sigma_\beta,\sigma_\gamma}={}_sR^{+}_{\omega,l,-\sigma_\gamma,\sigma_\beta,\sigma_\gamma}(r)$.
For them the \eqref{PolynomCond:a} gives an infinite series of
pure imaginary equidistant spectrum of frequencies
\ben
{}_s\omega^{+}_{N\sigma_\beta\sigma_\gamma}=\sigma_\gamma{\frac {i} 2}(N+1+\sigma_\beta s), \,\,\,N=0,1,2,\dots
\la{omegaNp}
\een

Making use of \eqref{munu:b} we obtain for the parameters in \eqref{RWparameters:a}
\ben
{}_s\mu^{+}_{N E\, \sigma_\beta\sigma_\gamma}=E +(1+\sigma_\beta s)^2+(1+2\sigma_\beta s)N,
\,\,\,N=0,1,2,\dots,
\la{muNp}
\een
with $E=l(l+1)$, where $l=|s|,|s|+1,|s|+2,\dots$ is an {\em integer}.

B. In the second subcase we consider the solutions ${}_sR^{-}_{\omega,l,\sigma_\alpha,\sigma_\beta,\sigma_\gamma}(r)$
with $\sigma_\alpha=-\sigma_\beta$. We denote them as ${}_sR^{-}_{\omega,l,\sigma_\beta,\sigma_\gamma}={}_sR^{-}_{\omega,l,-\sigma_\beta,\sigma_\beta,\sigma_\gamma}(r)$.
For them the \eqref{PolynomCond:a} gives an infinite series of
pure imaginary equidistant spectrum of frequencies
\ben
{}_s\omega^{-}_{N\sigma_\beta\sigma_\gamma}=\sigma_\beta{\frac {i} 2}(N+1+\sigma_\gamma s), \,\,\,N=0,1,2,\dots
\la{omegaNm}
\een

Making use of \eqref{munu:b} we obtain for the parameters in \eqref{RWparameters:b}
\ben
{}_s\mu^{-}_{N E\,\sigma_\beta\sigma_\gamma}= E + \sigma_\gamma s N +(N+1+\sigma_\gamma s)^2,
\,\,\,N=0,1,2,\dots,
\la{muNm}
\een
with $E=l(l+1)$, where $l=|s|,|s|+1,|s|+2,\dots$ is an {\em integer}.

If $E=l(l+1)$, we will have regular solutions to the angular equation -- the spin-weighted spherical functions.
It is easy to check that in this case the $\Delta_{N+1}$-condition can not be satisfied for $N=2$ and for $N\geq 4$.
In contrast, for $N=1,3$ we have several new polynomial solutions to the TRE:

1. The $\Delta_{N+1}$-condition is fulfilled for the solutions ${}_sR^{+}_{\omega,l,\sigma_\alpha,\sigma_\beta,\sigma_\gamma}(r)$ when $N=1$ and:

a) $s=2$, $l=-2,1,\,\sigma_{\alpha}=\mp 1,\,\sigma_{\beta}=-1,\,\sigma_{\gamma}=\pm 1$;

b) $s=-2$, $l=-2,1,\,\sigma_{\alpha}=\mp 1,\,\sigma_{\beta}=1,\,\sigma_{\gamma}=\pm 1$.

2. The $\Delta_{N+1}$-condition is fulfilled for the solutions ${}_sR^{-}_{\omega,l,\sigma_\alpha,\sigma_\beta,\sigma_\gamma}(r)$ when $N=1$ and:

a) $s=1$, $l=-1,0,\,\sigma_{\alpha}=1,\,\sigma_{\beta}=\pm 1,\,\sigma_{\gamma}=-1$;

b) $s=-1$, $l=-1,0,\,\sigma_{\alpha}=-1,\,\sigma_{\beta}=\pm 1,\,\sigma_{\gamma}=1$;

c) $s=2$, $l=-2,1,\,\sigma_{\alpha}=1,\,\sigma_{\beta}=\pm 1,\,\sigma_{\gamma}=-1$;

d) $s=-2$, $l=-2,1,\,\sigma_{\alpha}=-1,\,\sigma_{\beta}=\pm1,\,\sigma_{\gamma}=1$.

Since these solutions do not satisfy the requirement $l\geq |s|$ for the solutions to angular equation,
they seem to be incidental. We include them for completeness of our list of polynomial solutions to the radial RWE.

3. The $\Delta_{N+1}$-condition is fulfilled for the solutions ${}_sR^{+}_{\omega,l,\sigma_\alpha,\sigma_\beta,\sigma_\gamma}(r)$
when  $N=3$ and:

a) $s=1$, $l=-2,1,\,\sigma_{\alpha}=\pm 1,\,\sigma_{\beta}=- 1,\,\sigma_{\gamma}= \mp 1$;

b) $s=-1$, $l=-2,1,\,\sigma_{\alpha}=\mp 1,\,\sigma_{\beta}= 1,\,\sigma_{\gamma}=\pm 1$.

The solutions with $N=3$ and $l=-2$ do not satisfy the additional condition  $l\geq |s|$,
but for $l=1$ we obtain two new algebraically special solutions to the RWE of electromagnetic type
with $\omega^{+}_{\sigma_\gamma}=\pm i3/2$,
which are not described in the known to us literature.

The situation with the polynomial solutions to the RWE will be drastically changed if one rejects the regularity
requirement on the solutions to the angular equation. As a result the condition $E=l(l+1)$ will be no more valid.
Then in the case of polynomial solutions of second class one can use the \eqref{omegaNp} and \eqref{muNp},
or the \eqref{omegaNm} and \eqref{muNm} in the $\Delta_{N+1}$-condition. Thus one obtains
an algebraic equation of degree $(N+1)$ for the constant $E$ with solutions
$E={}_sE^{\pm}_{N,n,\sigma_\alpha,\sigma_\beta,\sigma_\gamma}$ where $n=0,\dots, N$.

The infinite series of polynomial solutions of second class ${}_sR^{\pm}_{N,n,\sigma_\alpha,\sigma_\beta,\sigma_\gamma}(r)$,
entering the expression
${}_s\Phi_{N,n,\sigma_\alpha,\sigma_\beta,\sigma_\gamma}(t,r)=e^{-i\omega_{N,n,\sigma_\alpha,\sigma_\beta,\sigma_\gamma} t} {}_sR^{\pm}_{N,n,\sigma_\alpha,\sigma_\beta,\sigma_\gamma}(r)$,
describes non-oscillating exponentially decaying in time linear perturbations of SBH,
when $\Im\left({}_s\omega^{\pm}_{N\sigma_\beta\sigma_\gamma} \right)<0$.
Obviously, the infinite set of these solutions for $N=0,1,2,\dots$
presents a complete basis for Laplace series expansion in time $t$
for linear perturbations of more general form. The angular factor
${}_sS^{\pm}_{N,n,\sigma_\alpha,\sigma_\beta,\sigma_\gamma}(\theta)$
of the corresponding kernel
\ben
{}_s{\mathcal K}_{N,n,m,\sigma_\alpha,\sigma_\beta,\sigma_\gamma}(t,r,\theta,\varphi)=
e^{-i\omega_{N,n,\sigma_\alpha,\sigma_\beta,\sigma_\gamma} t} e^{im\varphi}\,
{}_sS^{\pm}_{N,n,\sigma_\alpha,\sigma_\beta,\sigma_\gamma}(\theta)\,
{}_sR^{\pm}_{N,n,\sigma_\alpha,\sigma_\beta,\sigma_\gamma}(r)
\la{DiscreteKernelRWE}
\een
will be singular at least at one of the poles $\theta=0$, $\theta=\pi$,
or simultaneously at both of them.
This causes a strong anisotropy of the emission of the corresponding waves,
i.e.,  a collimation of the emission around the poles.
This kernel can be used for representation of this type of solution to the full RWE in the form
\ben
{}_s\Phi(t, r, \theta, \varphi)=\sum_{N,n,m,\sigma_\alpha,\sigma_\beta,\sigma_\gamma}
{}_sA_{N,n,m,\sigma_\alpha,\sigma_\beta,\sigma_\gamma}\,\,
{}_s{\mathcal K}_{N,n,m,\sigma_\alpha,\sigma_\beta,\sigma_\gamma}(t,r,\theta,\varphi).
\la{DiscreteSolRWE}
\een
The regularity of these solutions and the convergence of discrete sum \eqref{DiscreteSolRWE}
depends on the choice of the amplitudes $A_{N,n,m,\sigma_\alpha,\sigma_\beta,\sigma_\gamma}$.

One can find analogous finite set of polynomial solutions of the first class to the radial RWE,
which correspond to singular angular part of the total linear perturbation, too.

We will not discuss here these $r$-polynomial-$\theta$-singular solutions in detail,
since it is more convenient to obtain the corresponding physical information in terms of
perturbations of Weyl scalars using TME in the limit $a\to 0$. This will be done in the next Sections.

\section{Exact Solutions to the Teukolsky Radial Equation in Terms of the Confluent Heun's functions}
The explicit form of TRE
\ben
\la{TRE}
{\frac {d^{2}R_{\omega,E,m}}{d{r}^{2}}} + (1+s)  \left( {\frac{1}{r-{\it r_{+}}}}+
{\frac{1}{r-{\it r_{-}}}} \right){\frac {dR_{\omega,E,m}}{dr}} + \nonumber\\                                                                                                 +\left( {\frac { \Big( \omega\, \left( {a}^{2}+{r}^{2} \right) -am \Big) ^{2}}{ \left( r-r_{+} \right)  \left( r-r_{-} \right) }}-                             is \left( {\frac{1}{r-{\it r_{+}}}}+ {\frac{1}{r-{\it r_{-}}}} \right)  \Big( \omega\, \left( {a}^{2}+{r}^{2} \right) -am \Big)- \right.\nonumber\\
\left.-E+s(s+1)-{a}^{2}{\omega}^{2}+2m\,a\omega + 4\,i s \omega r  \vphantom{\frac{\Big( a^2\Big)}{\big( a^2 \big)}}\!\right)
{\frac {R_{\omega,E,m}} {( r-r_{+})( r-r_{-})}}
\een
shows that it has three singular points: $r=r_{\pm}$ and $r=\infty$.
In the present article we consider only the {\em non-extremal} Kerr metric with {\em real} $r_{+}>r_{-}\geq 0$.
Then precisely as in the case of RWE,
the first two are regular singular points (the Cauchy horizon $r=r_{-}$
and the event horizon $r=r_{+}$),
and the third one (the physical infinity $r=\infty$) is irregular singular point.
The symmetry of the \eqref{TRE} under the interchange $r_{+}\leftrightarrows r_{-}$ is obvious.

As well as in the case of SBH, the algebraic invariants of Riemann tensor are not able to indicate the horizons of KBH
and one usually considers them as a pure coordinate singularities of the metric in Boyer-Lindquist coordinates.
In contrast, the circle $r=0, \theta = \pi/2$ is a singularity of the algebraic invariants of Riemann tensor \cite{Chandra}.

It is not too hard to find differential invariants of the Riemann tensor,
which are able to distinguish both horizons $r=r_{\pm}$ and the ergosphere $g_{tt}=0$.
Indeed, let us consider the following algebraic invariants of Weyl tensor ${\cal W}_{ijkl}$:
$I_{1}={\frac 1 {48}}{\cal W}_{ijkl}{\cal W}^{ijkl}$ -- the density of the Euler characteristic class,
and
$I_{2}={\frac 1 {48}}{\cal W}_{ijkl}\,{}^*{\cal W}^{ijkl}$ -- the density of the Chern-Pontryagin characteristic class \cite{char_classes}.
Let us put $(I_{1}-iI_{2})^{1/2}=\lambda=|\lambda|\exp(i\psi)$. Then $r=\left({\frac M{|\lambda|}}\right)^{1/6}\cos(\psi/6)$ and
$\rho=\left({\frac M{|\lambda|}}\right)^{1/6}\cos(\psi/6)^{-1}$ are obviously invariants of Weyl tensor
-- non-algebraic and non-differential ones.
In Boyer-Lindquist coordinates one obtains $\rho=r+{\frac{a^2}r}\cos\theta$ and $g_{tt}=1-2M/\rho$.
The differential invariants of first order
\begin{subequations}
\label{KBH_D_inv:ab}
\ben
DI_{1}=-\left(\nabla \ln r\right)^2={\frac{1}{r\rho}}\left(1-{\frac{2M}{r}}+{\frac{a^2}{r^2}}\right),\label{KBH_D_inv:a}\\
DI_{2}=\left(\nabla \ln \rho\right)^2-\left(\nabla \ln r\right)^2=
{\frac 4 {\rho^2}}\left({\frac \rho r}-1\right)\left(1-{\frac{2M}\rho}\right)\label{KBH_D_inv:b}
\een
\end{subequations}
indicate the two KBH horizons, the ergo-sphere and some other geometrical objects in Kerr space-time.
The very horizons are singularities of the same kind in the equation \eqref{TRE}.
The relations in \eqref{KBH_D_inv:ab} generalize the equation \eqref{diffInvSBH},
which is a limiting case of equation \eqref{KBH_D_inv:a} for $a\to 0$, when $\rho\to r$.
In this limit the invariant in equation \eqref{KBH_D_inv:b} becomes trivial: $DI_{2}\to 0$.
Hence, at this point we have a complete analogy between RWE and TRE.

The analytical study of the solutions to the TRE and TAE was started in \cite{Leaver}
and continued by different approximate methods (see in \cite{QNM, KerrQNM_analytic}) without utilizing of Heun functions.
Using the confluent Heun function one can write down 16 exact local Frobenius type solutions to the TRE (\ref{TRE}) in the form:
\ben
{}_sR^\pm_{\omega,E,m,\sigma_\alpha,\sigma_\beta,\sigma_\gamma}(r;r_{+},r_{-})\Delta^{s/2}\!=\!e^{\sigma_\alpha{\frac{\alpha_{{}_\pm} z_{{}_\pm}} 2}} z_{{}_\pm}^{\sigma_\beta{\frac{\beta_{{}_\pm}} 2}} z_{{}_\mp}^{\sigma_\gamma{\frac{\gamma_{{}_\pm}} 2}}
\text{HeunC}({\sigma_\alpha\alpha_{{}_\pm},\sigma_\beta\beta_{{}_\pm},\sigma_\gamma\gamma_{{}_\pm},\delta_{{}_\pm},\eta_{{}_\pm},z_{{}_\pm}}),
\la{TRE16local}\een
which is very similar to the form of the equation \eqref{RWE16local}.  Now\footnote{Note that the notations $z_{\pm}$
in the \eqref{TREparameters:f} are consistent with
the corresponding ones for the case of RWE -- \eqref{RWparameters:d}. They are based on the limits $z_{\pm}\to \pm \infty$ for $r\to\infty$.
Their relation with the notations of the parameters of KBH $r_{\pm}$ is illustrated by the equations $z_{\pm}(r_{\mp};r_{+},r_{-})=0$.
The labels $\pm$ in the notation $R^{\pm}$ -- \eqref{TRE16local} are related to the labels of their arguments  $z_{\pm}$,
not with the labels of the parameters $r_{\pm}$.}
\begin{subequations}\label{TREparameters:abcdef}
\begin{align}
\alpha_{{}_{+}}\!&=\!{}_{s}\alpha_{\omega,E,m}(r_{+},r_{-})\!=\!2i\omega(r_{+}-r_{-})\!=i p\, {{\omega}/{\Omega}_a}, \label{TREparameters:a}\\
\beta_{{}_{+}}&=\!{}_{s}\beta_{\omega,E,m}(r_{+},r_{-})\!=s + 2 i \left(m-\omega/\Omega_{-}\right)/p,\label{TREparameters:b}\\
\gamma_{{}_{+}}&=\!{}_{s}\gamma_{\omega,E,m}(r_{+},r_{-})\!=s - 2 i \left(m-\omega/\Omega_{+}\right)/p, \label{TREparameters:c}\\
\delta_{{}_{+}}&=\!{}_{s}\delta_{\omega,E,m}(r_{+},r_{-})\!=\alpha_{+}\left(s-i\omega(r_{+}+r_{-})\right)\!=
\alpha_{+}\left(s-i\omega/\Omega_{g}\right),\label{TREparameters:d}\\
\eta_{{}_{+}}\!&=\!{}_{s}\eta_{\omega,E,m}(r_{+},r_{-})\!=\!-E+s^2\!+m^2\!+
{\frac{2m^2\Omega_a^2-\omega^2}{p^2\Omega_a^2}}-{\frac{(2m\Omega_a-\omega)^2}{p^2\Omega_g^2}}
-{\frac 1 2}\left({s\!-\!i\,\frac{\omega\,\Omega_{+}}{\Omega_a\Omega_g}}\right)^2;\label{TREparameters:e}\\
z_{+}\!&=\!z_{+}(r;r_{+},r_{-})\!=\!{\frac{r\!-\!r_{-}}{r_{+}\!-\!r_{-}}},\,z_{-}\!=\!z_{-}(r;r_{+},r_{-})\!=\!{\frac{r_{+}\!-\!r}{r_{+}\!-\!r_{-}}};\,
z_{+}\!+\!z_{-}\!=\!1,\,z_{+}z_{-}\!=\!{\frac{-\Delta}{(r_{+}\!-\!r_{-})^2}}.\label{TREparameters:f}
\end{align}
\end{subequations}

The discrete parameters $\sigma_\alpha,\sigma_\beta,\sigma_\gamma$ are the same as in \eqref{RWE16local}.
In equations \eqref{TREparameters:a}-\eqref{TREparameters:e} we are using the following quantities:
the angular velocity of the event horizon $\Omega_{+}={\sqrt{r_{-}/r_{+}}}\big/\left(r_{+}+r_{-}\right)$,
the angular velocity of the Cauchi horizon $\Omega_{-}={\sqrt{r_{+}/r_{-}}}\big/\left(r_{+}+r_{-}\right)$,
the arithmetically-averaged angular velocity $\Omega_a=\left(\Omega_{+}+\Omega_{-}\right)/2=1/(2a)$,
the geometrically-averaged angular velocity $\Omega_g=\sqrt{\Omega_{+}\Omega_{-}}=1/(2M)$, and the new dimensionless parameter
$p=\sqrt{r_{+}/r_{-}}-\sqrt{r_{-}/r_{+}}=\sqrt{\Omega_{-}/\Omega_{+}}-\sqrt{\Omega_{+}/\Omega_{-}}\in (0,\infty)$.
Note that the inverse relation
$r_{\pm}={\sqrt{\Omega_{\mp}/\Omega_{\pm}}}\big/\left(\Omega_{+}+\Omega_{-}\right)$
permits us to replace $r_\pm$ with $\Omega_\pm$ wherever it is necessary,
thus making transparent the duality of the parameters $r_\pm$ and $\Omega_\pm$,
as well as the behavior of the above quantities under interchange of the two horizons:
$r_{+}\leftrightarrows r_{-}$ $\Rightarrow$  $\Omega_{+}\leftrightarrows \Omega_{-}$,
$p\mapsto -p$, $\Omega_{a,g}\mapsto\Omega_{a,g}$ -- invariant.

The parameters $\alpha_{{}_{-}}, \beta_{{}_{-}}, \gamma_{{}_{-}}, \delta_{{}_{-}}, \eta_{{}_{-}}$
can be obtained  by interchanging the places of the two horizons: $r_{+}\leftrightarrows r_{-}$
in \eqref{TREparameters:a} -- \eqref{TREparameters:e}.
This procedure may be substantiated using the known properties of the confluent Heun function under
changes of parameters \cite{Heun}. One can check directly that in this way we obtain indeed solutions of \eqref{TRE}.

According to equation \eqref{TRE16local} and equation \eqref{TREparameters:f} the behavior of the solutions
${}_sR^\pm_{\omega,E,m,\sigma_\alpha,\sigma_\beta,\sigma_\gamma}(r;r_{+},r_{-})$
around the corresponding singular points $z_{\pm}=0=z_{\pm}(r_{\mp};r_{+},r_{-})$ is defined by the dominant factor
$z_{{}_\pm}^{\sigma_\beta \beta_{\pm}/2}$.
All other factors in equation \eqref{TRE16local} are regular around these points.
The same solutions
are in general singular around the corresponding singular points $z_{\pm}=1=z_{\pm}(r_{\pm};r_{+},r_{-}$.
Hence, at this point we have a complete analogy with the Regge-Wheeler case, including our notations.

Only two of the sixteen solutions  (\ref{TRE16local}) are linearly independent.
For different purposes one can use different pairs of independent local solutions.

Using the known asymptotic expansion of the confluent Heun function \cite{Heun}
we obtain, as a generalization of \eqref{RWEinf}, two asymptotic solutions of Tom\`{e} type. These are
local solutions of TRE around its irregular singular point $|r|=\infty$  in the complex plane $\mathbb{C}_r$:
\ben
{}_sR_{\omega,E,m,\sigma_\alpha,\sigma_\beta,\sigma_\gamma}^{\pm\infty}(r;r_{+},r_{-})\sim
{{e^{ i\sigma_{\!\alpha}\,\omega \big(r + (r_{+}+r_{-})\ln r\big)}}}
\sum_{j\geq 0}a_{j}\left(\pm{{r_{+}-r_{-}}\over{r}}\right)^{j+1+(1+\sigma_\alpha)s},\,\,\,\,a_{0}=1.
\la{TMEinf}\een
The notation $\pm\infty$ in \eqref{TMEinf} denotes the two directions: $r\to +\infty$ and  $r\to -\infty$ on the real $r$-axes
for approaching the irregular singular point $|r|=\infty$  in the complex plane $\mathbb{C}_r$.
For the coefficients $a_{j}=a_{j,\omega,E,m,\sigma_\alpha,\sigma_\beta,\sigma_\gamma}$  one has a recurrence relation \cite{Heun}
which shows that they increase together with the integer $j$. Hence, the asymptotic series (\ref{TMEinf}) is a divergent one.

As seen from \eqref{TREparameters:abcdef}, ${}_sR^{-}_{\omega,E,m,\sigma_\alpha,\sigma_\beta,\sigma_\gamma}(r;r_{+},r_{-})=
{}_sR^{+}_{\omega,E,m,\sigma_\alpha,\sigma_\beta,\sigma_\gamma}(r;r_{-},r_{+})$.
Hence, one can introduce a new parity property of the solutions and construct a symmetric and anti-symmetric
(with respect to the interchange $r_{+} \rightleftarrows r_{-}$) solutions of TRE:
\ben
\la{aSymTRE}
{}_sR^{SYM}_{\omega,E,m,\sigma_\alpha,\sigma_\beta,\sigma_\gamma}(r;r_{+},r_{-})={\frac 1 2}
\left({}_sR^{+}_{\omega,E,m,\sigma_\alpha,\sigma_\beta,\sigma_\gamma}(r;r_{+},r_{-})+
{}_sR^{-}_{\omega,E,m,\sigma_\alpha,\sigma_\beta,\sigma_\gamma}(r;r_{+},r_{-})\right),\\
{}_sR^{ASYM}_{\omega,E,m,\sigma_\alpha,\sigma_\beta,\sigma_\gamma}(r;r_{+},r_{-})={\frac 1 2}
\left({}_sR^{+}_{\omega,E,m,\sigma_\alpha,\sigma_\beta,\sigma_\gamma}(r;r_{+},r_{-})-
{}_sR^{-}_{\omega,E,m,\sigma_\alpha,\sigma_\beta,\sigma_\gamma}(r;r_{+},r_{-})\right).
\nonumber
\een
Clearly, these solutions are singular at both horizons in the general case,
but when one considers the two-singular-point boundary problem \cite{Heun}
on the interval $[r_{-},r_{+}]$  in KBH interior,
the solutions \eqref{aSymTRE} may be regular at one, or at the both ends
for some values of the separation constants $\omega$ and $E$.
Since this boundary problem is still not studied,
at present we are not able to make more definite statements about this case.

\section{Classification of the solutions to TRE based on the  $\delta$-condition}

For TRE  the $\delta$-condition reads:
\ben
{}_s\omega^\pm_{m,\sigma_\alpha,\sigma_\beta,\sigma_\gamma}\, {\cal L}^{\pm}_{\sigma_\alpha,\sigma_\beta,\sigma_\gamma}=
\Omega_g\left({\cal M}^{\pm}_{m,\sigma_\alpha,\sigma_\beta,\sigma_\gamma} +
i\, {}_s{\cal N}^\pm_{\sigma_\alpha,\sigma_\beta,\sigma_\gamma}\right),
\la{TRE_delta_cond}
\een
where $${\cal L}^{\pm}_{\sigma_\alpha,\sigma_\beta,\sigma_\gamma}\!=\!
{\frac{\sigma_\beta \Omega_{\pm}\!-\!\sigma_\gamma\Omega_{\mp}}{\Omega_{\pm}\!-\!\Omega_{\mp}}}\!-\!\sigma_\alpha,\,\,
{\cal M}^{\pm}_{m,\sigma_\alpha,\sigma_\beta,\sigma_\gamma}\!=\!m(\sigma_\beta\!-\!\sigma_\gamma){\frac{\Omega_g}{\Omega_{\pm}\!-\!\Omega_{\mp}}},\,\,
{}_s{\cal N}^\pm_{\sigma_\alpha,\sigma_\beta,\sigma_\gamma}\!=\!N\!+\!1\!+\!\left(\sigma_\alpha\!+\!{\frac{\sigma_\beta\!+\!\sigma_\gamma} 2}\right)s.$$

The calculation of the values of the coefficients in equation (\ref{TRE_delta_cond})
yields two very different cases:

1. In the first case from relations  ${\cal L}^{+}_{\pm,\pm,\pm}={\cal L}^{-}_{\pm,\pm,\pm}=0$
we see that one is not able to fix the frequencies
${}_s\omega^{+}_{m,\pm,\pm,\pm}$ and ${}_s\omega^{-}_{m,\pm,\pm,\pm}$.
Instead, choosing $\sigma_\alpha=\sigma_\beta = \sigma_\gamma =-\text{sign} (s)\equiv -\sigma$
and using (\ref{TRE_delta_cond}) one fixes the degree of the polynomial
$\Delta_{N+1}$-condition in the form
\ben
{}_sN+1= 2|s|.
\la{sN_Kerr}
\een

2. In the second case the coefficients ${\cal L}^{\pm}_{\sigma_\alpha,\sigma_\beta,\sigma_\gamma}$ are not zero and
one can fix the values of the frequencies ${}_s\omega^\pm_{m,\sigma_\alpha,\sigma_\beta,\sigma_\gamma}$
from equation (\ref{TRE_delta_cond}). Thus one obtains two different types of exact {\em equidistant} spectra:

a) For ${\cal L}^{+}_{\mp,\pm,\pm}={\cal L}^{-}_{\mp,\pm,\pm}=\pm 2$, ${\cal M}^{+}_{\mp,\pm,\pm}={\cal M}^{-}_{\mp,\pm,\pm}=0$ and
${\cal N}^{+}_{\mp,\pm,\pm}={\cal N}^{-}_{\mp,\pm,\pm}=(N+1)$ the $\delta$-condition (\ref{TRE_delta_cond})
produces the pure imaginary equidistant frequencies
\ben
{}_s\omega^{+}_{N,m,\mp,\pm,\pm}={}_s\omega^{-}_{N,m,\mp,\pm,\pm}=\pm i\,{\frac {N+1} {4M}},\,\,\, N\geq 0\,\,\,\text{-- integer}.
\la{Im_omega_N_Kerr}
\een
Note that these frequencies do not depend on the spin-weight  $s$ and azimuthal number $m$, nor on the rotation parameter $a$,
i.e.,  this spectrum is not influenced  by the rotation of the waves and the rotation of Kerr metric.
The frequencies (\ref{Im_omega_N_Kerr}) are defined only by the monopole term in multipole expansion of the metric.

b) For all other cases the coefficients ${\cal L}^{\pm}_{\sigma_\alpha,\sigma_\beta,\sigma_\gamma}$,                                                       and ${\cal M}^{\pm}_{\sigma_\alpha,\sigma_\beta,\sigma_\gamma}$
are not constant and one obtains the following two similar double-equidistant spectra of frequencies:

\begin{subequations}\label{omega_N:ab}
\ben
{}_s\omega^{+}_{N,m,\mp,\mp,\pm}={}_s\omega^{-}_{N,m,\mp,\pm,\mp}
=m\Omega_{+}\pm{\frac{i}{4M}}\left(1-{\frac{r_{-}}{r_{+}}}\right)(N+1 \mp s),
\,\,\, N\geq 0,\,\,\,m\,\,\,\text{-- integers}, \label{omega_N:a}\\
{}_s\omega^{+}_{N,m,\pm,\mp,\pm}={}_s\omega^{-}_{N,m,\pm,\pm,\mp}
=m\Omega_{-}\pm{\frac{i}{4M}}\left({\frac{r_{+}}{r_{-}}}-1\right)(N+1 \pm s),
\,\,\, N\geq 0,\,\,\,m\,\,\,\text{-- integers}. \label{omega_N:b}
\een
\end{subequations}

\section{Polynomial Solutions to the TRE}

The $\delta$-condition is not sufficient to ensure polynomial character of the solutions,
but it yields the basic classification of the solutions, described in previous Section 5.
In accord with it one obtains two classes of polynomial solutions to the TRE,
imposing in addition the $\Delta_{N+1}$-condition \eqref{Delta}.

\subsection{First Class of Polynomial Solutions to TRE:}

The solutions of this class correspond to the first case in the Section 5 and obey
the equation (\ref{sN_Kerr}).
The inequality  ${}_sN=2|s|-1 \geq 0$ excludes the existence of scalar perturbations ($|s|=0$)
of first polynomial class, just as in the case of RWE.

\subsubsection{The General Case:}
For brevity we denote the solutions ${}_sR^{\pm}_{\omega,E,m,-\sigma,-\sigma,-\sigma}(r;r_{+},r_{-})$
as ${}_sR^{\pm}_{\omega,E,m}(r;r_{+},r_{-})$.
For them the parameter $\mu$ takes the values
$\mu={}_s\mu_{\omega,k,m}^{\pm}(r_{+},r_{-}),\,\,\, k=1,\dots, 2|s|$ --
the solutions of the algebraic equation \eqref{Delta}, which now takes the form: $\Delta^{\pm}_{2|s|}(\mu)=0$.
Its degree is $2|s|=1, 2, 3$, or $4$, depending on the spin of the perturbations $|s|=1/2, 1, 3/2, 2$.
Making use of \eqref{munu:b}, and \eqref{TREparameters:a}-\eqref{TREparameters:e},
we obtain for the separation constant $E={}_sE_{\omega,k,m}^{\pm}(r_{+},r_{-})$,
$k\!=\!1,\dots, 2|s|$ the expressions
\ben
{}_sE_{\omega,k,m}^{\pm}(r_{+},r_{-})\!=\!{}_s\mu_{\omega,k,m}^{\pm}(r_{+},r_{-})
+|s|(|s|-1)-a\omega(a\omega-2m)+2i\sigma(2|s|-1)\omega r_{\mp}, \label{E_first_class}
\een

Applying the explicit expressions for the roots ${}_s\mu_{\omega,k,m}^{\pm}(r_{+},r_{-})$,
we obtain:
\ben
{}_sE_{\omega,m}^{\pm}(r_{+},r_{-})\!=-a\omega(a\omega-2m)-{\frac 1 4}:\,\,\,\text{for}\,\,\,|s|={\frac 1 2}.
\la{E_first_class_1/2}
\een
\ben
{}_sE_{\omega,k,m}^{\pm}(r_{+},r_{-})\!=-a\omega(a\omega-2m)-2(-1)^k\sqrt{a\omega(a\omega-m)}:\,\,\,\text{for}\,\,\,|s|=1,\,k=1,2.
\la{E_first_class_1}
\een

For the gravitational waves ($|s|=2$) one has  to find the quantities
${}_s\mu_{\omega,k,m}^{\pm}(r_{+},r_{-})$ solving algebraic equation
of fourth degree $\Delta^{\pm}_{4}(\mu)=0$.
The explicit form of its roots is too complicated and not necessary for the purposes of present article.
It is more instructive to demonstrate here the result,
obtained using the Taylor series expansion of the solutions ${}_s\mu_{\omega,k,m}^{\pm}(r_{+},r_{-})$
around the zero frequency $\omega=0$.

Thus we obtain  for $|s|=2,\,\,k=1,2$ the eight  values:
\ben\label{mu:1_2}
{}_sE_{\omega,k,m}^{\pm}\!=\!2\!-\!4\left(m\!-\!i(-1)^k {\frac{3M}{2a}}\right)a\omega\!+\!
6\!\left(\!m^2\!+\!i(-1)^k2 m\left((m^2\!-\!1){\frac a M}\!+\!{\frac {2M} {a}}\right)\!+\!{\frac{3M^2}{a^2}}\!-\!{\frac{7}{6}}\right)(a\omega)^2
\!+\\
+{\cal{O}}_{3}(a\omega).\nonumber
\een
For $|s|=2,\,\,m\neq 0,\,\,k=3,4$ we have another eight values:
\ben\label{mu:3_4}
{}_sE_{\omega,k,m}^{\pm}=
i\,(-1)^k 4 \sqrt{ma\omega}\Bigg(1+i\,3\left(1+\left({\frac{3M^2}{8a^2}}-{\frac{2}{3}}\right){\frac{1}{m^2}}\right)ma\omega
+\!{\cal{O}}_{2}(a\omega)\Bigg)+\hskip 2.7truecm\nonumber\\
+8 ma \omega-6\left(1+\left({\frac{3M^2}{a^2}}-{\frac{5}{6}}\right){\frac{1}{m^2}}\right)(ma\omega)^2+\!{\cal{O}}_{3}(a\omega).
\hskip .5truecm
\een

Clearly, these series describe two kinds of solutions with a completely different behavior around the origin $\omega=0$.
In particular, the series \eqref{mu:1_2} and \eqref{mu:3_4} have different limits: $2$ and $0$, correspondingly, when $\omega\to 0$.
For the solutions \eqref{mu:3_4} the origin $\omega=0$ is a branching point, etc.

The independence of the values of ${}_sE_{\omega,k,m}^{\pm}$ in \eqref{mu:1_2} and \eqref{mu:3_4} on the $\pm$ signs
is a result of the polynomial character of the solutions, i.e. of the regularity of the corresponding HeunC-factor
simultaneously on both horizons $r_{\pm}$.

For a complete solution of the problem one has to determine the frequency $\omega$. Hence,
one needs additional relation between the parameters $E$ and $\omega$.
Such relation may appear when one solves the TAE (See next Section 8.).

\subsubsection{The Special Case of Schwarzschild metric:}
For the special value of the parameter  $a=0$ we have $r_{-}=0$, $r_{+}=2M$.
This is the case of perturbations to non-rotating SBH, now described in terms of Weyl scalars.
For consistence with the description of SBH perturbations by RWE
here we have to use units in which $2M=1$ (See the Introduction.).
The parameters in the solution (\ref{TRE16local}) acquire the limiting values
\ben
\alpha_{+}=2i\omega, \beta_{+}=s, \gamma_{+}=s+2i\omega, \delta_{+}=2i\omega(s-i\omega), \eta_{+}=-E+{\frac{s^2}{2}};
\hskip 1.3truecm \nonumber\\
\alpha_{-}=-2i\omega, \beta_{-}=s+2i\omega, \gamma_{-}=s, \delta_{-}=-2i\omega(s-i\omega),
\eta_{-}=-E+{\frac{s^2}{2}}+2\omega^2+2is\omega.
\la{parameters_Schwarcschild}
\een
These differ from the values of the parameters (\ref{RWparameters:abcd})
for description of the perturbations to Schwarzschild metric in the RWE approach.

In the limit $a\to 0$ the equation \eqref{TRE_delta_cond} does not define the frequency $\omega$,
if $\sigma_\alpha=\mp\sigma_\beta=\pm\sigma_\gamma=-\sigma$,
because then one obtains ${\cal L}^{\pm}_{-\sigma,\pm\sigma,\mp\sigma}=0$.
If, in addition, $\sigma=\text{sign}(s)$, then the $\delta$-condition is fulfilled for the special
polynomial solutions of the first class, denoted as
${}_sR^{\pm}_{\omega,E,m}(r)={}_sR^{\pm}_{\omega,E,m,-\sigma,\pm\sigma,\mp\sigma}(r;1,0)$.
The equation \eqref{TRE_delta_cond} yields the relation ${}_sN=|s|-1 \geq 0$ --
precisely the same as in the case of polynomial solutions of first class to the RWE (See Section 3.0.1.).
Scalar perturbations of this type do not exist.

In case of integer spins $|s|=1,2$ the roots $\mu={}_s\mu_{\omega,k,m}^{\pm},\,\,\, k=1,\dots, |s|$
of the equations $\Delta^{\pm}_{|s|}(\mu)=0$,  \eqref{munu:b},
and \eqref{TREparameters:a}-(\ref{TREparameters:e}) with $r_{+}=1$, $r_{-}=0$ and $a=0$
produce the following simple expressions for $E={}_sE_{\omega,k,m}^{\pm}$, $k=1,\dots, |s|$:
\begin{eqnarray}
{}_sE_{\omega,m}^{\pm}&=&0:\,\,\,\text{for}\,\,\, |s|=1,\\
{}_sE_{\omega,k,m}^{\pm}&=&1-(-1)^k\sqrt{1-i6\sigma\omega}:\,\,\,\text{for}\,\,\, |s|=2,\,k=1,2.
\la{A_first_special}
\end{eqnarray}

For a complete solution of the problem, one needs an additional relation between the parameters $E$ and $\omega$.
Such relation may be found solving the TAE (See Section 8.).

\subsection{Second Class of Polynomial Solutions to TRE:}
According to section 5 the solutions of this class originate from the second case
of $\delta$-condition and fall into two subclasses: a) and b).
The complete definite frequencies
$\omega={}_s\omega^{\pm}_{N,m,\sigma_\alpha,\sigma_\beta,\sigma_\gamma}$
-- formulae \eqref{Im_omega_N_Kerr} and \eqref{omega_N:ab},
yield algebraic equations $\Delta^{\pm}_{N+1}(\mu)=0$ with
$(N+1)$ roots $\mu={}_s\mu^{\pm}_{N,n,m,\sigma_\alpha,\sigma_\beta,\sigma_\gamma}(r_{+},r_{-})$, $n=0, 1, \dots, N$.
It seems hard to derive explicit analytic expressions for these roots, but their numerical values can be easily obtained.
Using the values of ${}_s\mu^{\pm}_{N,n,m,\sigma_\alpha,\sigma_\beta,\sigma_\gamma}(r_{+},r_{-})$
and equations \eqref{munu:b}, \eqref{TREparameters:a}-\eqref{TREparameters:e}
we obtain complete {\em definite} values for the parameter
$E={}_sE_{N,n,m,\sigma_\alpha,\sigma_\beta,\sigma_\gamma}^{\pm}(r_{+},r_{-})$:

a) Putting $\sigma=\text{sign}(s)$, in the case of the frequencies \eqref{Im_omega_N_Kerr} we obtain
\ben\label{1E_second_class}
{}_sE_{N,n,m,\sigma,-\sigma,-\sigma}^{\pm}\!=\!{}_s\mu_{N,n,m,\sigma,-\sigma,-\sigma}^{\pm}+
|s|(|s|-1)+a\omega(3a\omega-2m)\!+4\,\omega^2r_{\mp}^2\!+\!2i\sigma\omega\big(2M|s|-r_{\pm}\big).
\een

b) In the case of the frequencies  \eqref{omega_N:a}, \eqref{omega_N:b} we have correspondingly:
\begin{subequations}\label{2E_second_class:ab}
\ben
{}_sE_{N,n,m,+,-,+}^{\pm}\!-{}_s\mu_{N,n,m,+,-,+}^{\pm}={}_sE_{N,n,m,-,+,-}^{\mp}\!-{}_s\mu_{N,n,m,+,-,+}^{\mp}=
\pm i\,{2(2a\omega-m)}/{p}-\hskip 1.95truecm\nonumber \\
-\Big(m^2+8m\left(1+M^2/a^2\right)a\omega+
\left(1+ 10\, r_{\mp}/r_{\pm}+9\,(r_{\mp}/r_{\pm})^2-4\,(r_{\mp}/r_{\pm})^3\right)\omega^2 r_{\pm}^2\Big)\big/{p^2},
\hskip 1.truecm\label{2E_second_class.a}\\
{}_sE_{N,n,m,+,+,-}^{\pm}\!-\!{}_s\mu_{N,n,m,+,+,-}^{\pm}\!=\!{}_sE_{N,n,m,-,-,+}^{\mp}\!-\!{}_s\mu_{N,n,m,-,-,+}^{\mp}\!=\!
\pm i\,2\Big(m\!+\!2a\omega\!\left(1\!-\!2M^2/a^2\right)\!\Big)\big/p-\hskip .2truecm\nonumber\\
-i2psa\omega-4\Big(m^2+2m\left(1\!-\!3M^2/a^2\right)a\omega-\left(1\!-\!5M^2/a^2\right)(a\omega)^2\Big)\big/p^2.
\hskip 1.truecm\label{2E_second_class.b}
\een
\end{subequations}

With $\omega$ and $E$ given by the equations \eqref{Im_omega_N_Kerr},
\eqref{omega_N:ab} and \eqref{1E_second_class}, \eqref{2E_second_class:ab}
we have no more free parameters in the problem at hand.
As a result the corresponding solutions to TAE are fixed unambiguously
by the designated group of equations,
obtained for the second class of polynomial solutions to TRE.

\section{Exact Solutions to the Teukolsky Angular Equation in Terms of the Confluent Heun's functions}

In terms of the variable $x=\cos\theta$ the TAE has three singular points. Two of them: $x_{-}=-1$ (i.e., $\theta_{-}=\pi$ -- "south pole")
and $x_{+}= 1$ (i.e., $\theta_{+}=0$ -- "north pole") are regular singular points. The third one $x_\infty=\infty$ is irregular singular point.
It is remarkable that introducing the notations
\ben
z_{+}=z_{+}(\theta)=\left(\cos(\theta/2)\right)^2,\,\,\, z_{-}=z_{-}(\theta)=\left(\sin(\theta/2)\right)^2, \,\,\,z_{+}+z_{-}=1;
\la{TAE_zpm}
\een
and
\ben
a_{\pm}=\pm 4a\omega,\,\,b_{\pm}=s\mp m,\,\,c_{\pm}=s\pm m,\,\,d_{\pm}=\pm 4s a\omega ,\,\,n_{\pm}={\frac{m^2+s^2}{2}}\mp 2s a\omega -a\omega^2-E.
\label{TAEparameters}
\een
we can write down the 16 local solutions of the TAE in the form
\ben
{}_sS_{\omega,E,m,\sigma_a,\sigma_b,\sigma_c}^{\pm}=
\!e^{\sigma_a{\frac{a_{{}_\pm} z_{{}_\pm}} 2}}
z_{{}_\pm}^{\sigma_b{\frac{b_{{}_\pm}} 2}} z_{{}_\mp}^{\sigma_c{\frac{c_{{}_\pm}} 2}}
\text{HeunC}(\sigma_a a_{{}_\pm},\sigma_b b_{{}_\pm},\sigma_c c_{{}_\pm},d_{{}_\pm},n_{{}_\pm},z_{{}_\pm}),
\la{TAE16local}
\een
which is very similar to the form of the \eqref{RWE16local} and \eqref{TRE16local}.

Note that according to the \eqref{TAE16local} the behavior of the solutions
${}_sS_{\omega,E,m,\sigma_a,\sigma_b,\sigma_c}^{\pm}$
around the corresponding singular points $z_{\pm}=0=z_{\pm}(\theta_{\mp})$ is defined by the dominant factor
$z_{{}_\pm}^{\sigma_b b_{{}_\pm}/2}$.
All other factors in \eqref{TAE16local} are regular around these points.
The same solutions
are in general singular around the corresponding singular points $z_{\pm}=1=z_{\pm}(\theta_{\pm})$.
Hence, at this point we have a complete analogy with the cases of RWE and TRE,
including our notations.

Only two of the sixteen solutions  (\ref{TAE16local}) are linearly independent.
For various purposes one can use different pairs of independent local solutions.

In the case of integer spin weights $s=0,\pm1,\pm2$ there exist an additional complication.
The confluent Heun functions $\text{HeunC}(\alpha,\beta,\gamma,\delta,\eta,z)$
are not defined when $\beta$ is a negative integer \cite{Heun}.
Therefore if we have a negative integer $\beta=\sigma_b b_{{}_\pm}<0$,
we must write down the corresponding solutions in the form
\ben
{}_sS_{\omega,E,m,\sigma_a,\sigma_b,\sigma_c}^{\pm}=
\!e^{\sigma_a{\frac{a_{{}_\pm} z_{{}_\pm}} 2}}
z_{{}_\pm}^{\sigma_b{\frac{b_{{}_\pm}} 2}} z_{{}_\mp}^{\sigma_c{\frac{c_{{}_\pm}} 2}}
\underline{\text{HeunC}}(\sigma_a a_{{}_\pm},\sigma_b b_{{}_\pm},\sigma_c c_{{}_\pm},d_{{}_\pm},n_{{}_\pm},z_{{}_\pm}).
\la{TAE16local_cc}
\een
For this purpose we define the {\em concomitant} confluent Heun function
\ben
\underline{\text{HeunC}}(\alpha,\beta,\gamma,\delta,\eta,z)=z^{-\beta}\text{HeunC}(\alpha,-\beta,\gamma,\delta,\eta,z)
\int{\frac{e^{-\alpha\zeta}\zeta^{\beta-1}(1-\zeta)^{-\gamma-1}}{\big(\text{HeunC}(\alpha,-\beta,\gamma,\delta,\eta,z)\big)^2}}d\zeta.
\la{HeunCcc}
\een
Note that this function is well defined for non-positive integer $\beta=\sigma_b b_{{}_\pm}\leq 0$,
together with the confluent function $z^{-\beta}\text{HeunC}(\alpha,-\beta,\gamma,\delta,\eta,z)$.
In this case the confluent function $z^{-\beta}\text{HeunC}(\alpha,-\beta,\gamma,\delta,\eta,z)$
represents the local regular solution around the singular point $z=0$
and the concomitant confluent function  $\underline{\text{HeunC}}(\alpha,\beta,\gamma,\delta,\eta,z)$
represents an independent local singular solution around this point. It can be shown that
for negative integer $\beta$ the concomitant confluent Heun function possess the form
\ben
\underline{\text{HeunC}}(\alpha,\beta,\gamma,\delta,\eta,z)=\sum_{n=1}^{|\beta|}{\frac{c_n}{z^n}} + h_1(z)+h_2(z)\ln(z),
\,\,\,{\text{all}}\,\,\,c_n\neq 0.
\la{ccH}
\een
Here $h_{1,2}(z)$ denote two functions of the complex variable $z$,
which are analytic in vicinity of the point $z=0$.
In the problem at hand $|\beta_\pm|=|s\mp m|$.
The logarithmic term presents in the concomitant confluent Heun function when $|\beta|=0$, too,
but then we have no poles in the solution \eqref{ccH}. Its form otherwise is similar to (\ref{ccH}).
One can reach the last results using general analytical methods, described, for example, in \cite{DE}.

\section{Regular solutions of the TAE}
Using \eqref{TAE_zpm}, \eqref{TAEparameters}, and \eqref{TAE16local} we write down
two independent solutions to the TAE in the form:
\begin{align}
{}_sS^{\pm\, reg}_{\omega,E,m}(\theta):= \left\{{\begin{matrix}
{e^{2a\omega\,\left(\cos{{\theta}\over{2}}\right)^2}\left(\cos{{\theta}\over{2}}\right)^{{\left|s-m\right|}}}
\left(\sin{{\theta }\over{2}}\right)^{{s+m}}\\
{e^{2a\omega\,\left(\sin{{\theta}\over{2}}\right)^2}\left(\sin{{\theta}\over{2}}\right)^{{\left|s+m\right|}}}
{\left(\cos{{\theta}\over{2}}\right)^{{s-m}}}
\end{matrix}}\right\} \times\hskip 4.truecm\nonumber\\
\mathrm{HeunC}\left(4\,a\omega,\,\left|\,s \mp m\,\right|, s \pm m
,\pm 4s\, a\omega ,(m^2+s^2)/2
\mp 2 s\,a\omega - a\omega^{2} - E,\, {\begin{matrix}{\left(\cos{{\theta}\over{2}}\right)^2}\\
{\left(\sin{{\theta }\over{2}}\right)^2}\end{matrix}}\,
\right).\hskip -.15truecm \la{RegAngSol}\end{align}

Then the physically obvious symmetry of the problem becomes
transparent: the solutions ${}_sS^{+\, reg}_{\omega,E,m}(\theta)$ and ${}_sS^{-\, reg}_{\omega,E,m}(\theta)$
in (\ref{RegAngSol}) interchange their places after the substitution $s \to -s,\,\,\, \theta \to
\pi-\theta$.

One will have solutions ${}_sS^{\,\,{}_{REG}}_{\omega,E,m}(\theta)$, regular at {\em both} poles, if and only if
${}_sS^{+\, reg}_{\omega,E,m}(\theta)=\text{const}\,\, {}_sS^{-\, reg}_{\omega,E,m}(\theta)$,
or, equivalently, if the Wronskian vanishes:
$W\left[{}_sS^{+\, reg}_{\omega,E,m}(\theta),{}_sS^{-\, reg}_{\omega,E,m}(\theta)\right]=0$.
This condition determines  $E=E(a\omega, m, s)$ in the form
$$E = (m^2+s^2)/2-a\omega^2+\varepsilon(a\omega, m, s).$$
Here $\varepsilon(a\omega, m, s)$ is a solution of the transcendental
equation
\begin{align}
{{\mathrm{HeunC}^\prime(4a\omega,|m+s|,s-m,-4sa\omega,+2sa\omega-\varepsilon,\left(\sin{{\theta
}\over{2}}\right)^2)}\over {\mathrm{HeunC}(4a\omega,|m+s|,
s-m,-4sa\omega, +2sa\omega-\varepsilon,\left(\sin{{\theta
}\over{2}}\right)^2)}}+\nonumber\\
{{\mathrm{HeunC}^\prime(4a\omega,|m-s|,s+m,+4sa\omega,-2sa\omega-\varepsilon,
\left(\cos{{\theta}\over{2}}\right)^2)}\over
{\mathrm{HeunC}(4a\omega,|m-s|,s+m,+4sa\omega,-2sa\omega-\varepsilon,
\left(\cos{{\theta}\over{2}}\right)^2)}}+\nonumber\\
4a\omega +{{|s-m|-(s-m)}\over{2\left(\cos{{\theta}\over{2}}\right)^2}}+
{{|s+m|-(s+m)}\over{2\left(\sin{{\theta}\over{2}}\right)^2}}=0,\la{E_regular_TAE}
\end{align}
valid simultaneously for all  values of $\theta\in (0,\pi)$. The symbol $\mathrm{HeunC}^\prime$
denotes the derivative of the confluent Heun function.

Since the Eq.(\ref{E_regular_TAE}) is fulfilled for $\pi\!-\!\theta\in\!
(0,\pi)$, if and only if it is fulfilled for $\theta\in\! (0,\pi)$,
its symmetry: $s\!\to\! -s$ is obvious. The properties of the
function $\mathrm{HeunC}$ \cite{Heun} yield the additional symmetry:
$m\!\to\! -m$ of the Eq.(\ref{E_regular_TAE}).

Let us consider the limit $a\omega\to 0$ of the regular solutions of TAE.
Since
\ben\text{HeunC}(0,\beta,\gamma,0,\eta,z)\!=\!
(1\!-\!z)^{\beta+\gamma+1+\sqrt{\beta^2+\gamma^2+1-4\eta}}\times \hskip 7.3truecm\nonumber\\
{}_2F_{1}\left(
{\frac{\beta\!+\!\gamma\!+\!1+\!\sqrt{\beta^2\!+\!\gamma^2\!+\!1\!-\!4\eta}}{2}},
{\frac{\beta\!+\!\gamma\!+\!1-\!\sqrt{\beta^2\!+\!\gamma^2\!+\!1\!-\!4\eta}}{2}};
\beta\!+\!1;z\right)\!,
\la{HC_F}
\een
in this limit the Heun functions in \eqref{RegAngSol} can be reduced to the Gauss hypergeometric ones. Thus
we obtain the solutions in the form
\begin{align}
{}_sS^{\pm\,reg}_{0,E,m}(\theta):= \left\{{\begin{matrix}
{\left(\cos{{\theta}\over{2}}\right)^{{\left|s-m\right|}}}
\left(\sin{{\theta }\over{2}}\right)^{{s+m}}\\
{\left(\sin{{\theta}\over{2}}\right)^{{\left|s+m\right|}}}
{\left(\cos{{\theta}\over{2}}\right)^{{s-m}}}
\end{matrix}}\right\}
{}_2F_1\!\left(a_{{}_{0,\pm}},b_{{}_{0,\pm}};c_{{}_{0,\pm}}; {\begin{matrix}{\left(\cos{{\theta}\over{2}}\right)^2}\\
{\left(\sin{{\theta }\over{2}}\right)^2}\end{matrix}}\,
\right).\hskip -.15truecm
\la{RegAngSol0}\end{align}
Here
\ben
a_{{}_{0,\pm}}\!=\!{\frac{|s\mp m|\!+\!s\pm m}{2}}\!+\!{\frac 1 2}\!+\!\sqrt{E+1/4},\,
b_{{}_{0,\pm}}\!=\!{\frac{|s\mp m|\!+\!s\pm m}{2}}\!+\!{\frac 1 2}\!-\!\sqrt{E+1/4},\,
c_{{}_{0,\pm}}\!=\!|s\mp m|\!+\!1.
\la{parametersTAE0}
\een
The condition (\ref{E_regular_TAE}) reduces, correspondingly to
\begin{align}
{
{{}_2F_1^{{}^\prime}\left(a_{{}_{0,+}},b_{{}_{0,+}};c_{{}_{0,+}};\left(\cos{{\theta}\over{2}}\right)^2\right)}
\over
{{}_2F_1
\left(a_{{}_{0,+}},b_{{}_{0,+}};c_{{}_{0,+}};\left(\cos{{\theta}\over{2}}\right)^2\right)}
}\!+\!
{
{{}_2F_1^{{}^\prime}\left(a_{{}_{0,-}},b_{{}_{0,-}};c_{{}_{0,-}};\left(\sin{{\theta}\over{2}}\right)^2\right)}
\over
{{}_2F_1
\left(a_{{}_{0,-}},b_{{}_{0,-}};c_{{}_{0,-}};\left(\sin{{\theta}\over{2}}\right)^2\right)}
}\!+\!
{{|s\!-\!m|\!-\!(s\!-\!m)}\over{2\left(\cos{{\theta}\over{2}}\right)^2}}+
{{|s\!+\!m|\!-\!(s\!+\!m)}\over{2\left(\sin{{\theta}\over{2}}\right)^2}}=0.
\la{0E_regular_TAE}
\end{align}

Using the well known properties of the Gauss hypergeometric function  ${}_2F_1$ one can derive
from \eqref{0E_regular_TAE} the spectrum $E(0;l,s,m)=l(l+1)$,
$l=l(s,m,n)=\max(|m|,|s|)+n,\,\, n=0,1,2,\dots$;
with real values of the separation constant $E(0;l,s,m)$.
The numerical analysis of the \eqref{E_regular_TAE},
written directly in terms of confluent Heun functions,
confirms this standard result for the  limit $a\omega\!=\!0$.
The corresponding regular confluent Heun functions in
\eqref{RegAngSol} in this case are reduced to Jacobi polynomials.

The solutions $E(a\omega;s,m,l)$ for small $a\omega$ have been studied
many times \cite{Teukolsky,angular} in the form of Taylor series
expansion $E(a\omega;s,m,l)\!=\!l(l\!+\!1)\!+\!\sum_{n=\!1}^\infty
E_n(a\omega)^n$ without the use of the Eq.(\ref{E_regular_TAE}).
A little bit surprisingly, the solutions (\ref{RegAngSol}) with $a\omega\neq 0$ , regular at both
poles, are not polynomial and can be represented as an infinite
series with respect to Jacobi polynomials.
Further we denote these regular solutions by ${}_sS^{\,{}_{REG}}_{\omega,l,m}(\theta)$.

\section{Classification of the solutions to TAE based on the  $\delta$-condition}

For solutions ${}_sS_{\omega,E,m,\sigma_a,\sigma_b,\sigma_c}^{\pm}$ (\ref{TAE16local}) to TAE  the $\delta$-condition reads:
\ben
0=\mp\, m{\frac{\sigma_b-\sigma_c}{2}} + N+1+\left(\sigma_a + {\frac{\sigma_b+\sigma_c}{2}} \right)s.
\la{TAE_delta_cond}
\een

Comparing this equation with the corresponding one for TRE
-- \eqref{TRE_delta_cond}, we see both essential differences and similarities.
For the coefficients in equation \eqref{TAE_delta_cond}, which are analogous to the ones in \eqref{TRE_delta_cond},
one obtains:
$${\cal L}^{\pm}_{\sigma_a,\sigma_b,\sigma_c}\equiv 0,\,\,
{\cal M}^{\pm}_{m,\sigma_a,\sigma_b,\sigma_c}=\mp \,m(\sigma_b\!-\!\sigma_c){\frac 1 2},\,\,
{}_s{\cal N}^\pm_{\sigma_a,\sigma_b,\sigma_c}=N\!+\!1\!+\!\left(\sigma_a\!+\!{\frac{\sigma_b\!+\!\sigma_c} 2}\right)s.$$

Hence:

i) The coefficients ${\cal L}^{\pm}_{\sigma_a,\sigma_b,\sigma_c}$ vanish identically,
in contrast to the coefficients  ${\cal L}^{\pm}$ in equation \eqref{TRE_delta_cond}.
Consequently, there are no cases in which the condition \eqref{TAE_delta_cond} can fix the frequencies $\omega$.

ii) The form of the coefficients ${\cal M}^{\pm}$ of both equations \eqref{TRE_delta_cond} and \eqref{TAE_delta_cond}
is the same only for $a^2/M^2=1/2$.

iii) The coefficients ${\cal N}^\pm$ of both equations are of the same form.

As a result we obtain two different cases:

1. For $\sigma_b=\sigma_c=\sigma_a=-\sigma$ equation \eqref{TAE_delta_cond} fixes the degree of the polynomial
$\Delta_{N+1}$-condition in the same form as equation \eqref{sN_Kerr}\footnote{The alternative case
$\sigma_b=\sigma_c=-\sigma_a$ leads to a non-interesting relation $N+1=0$.}:
\ben
{}_sN+1= 2|s|\geq 1\,\,\,\text{for}\,\,\,|s|\geq 1/2.
\la{sN_Kerr_A}
\een

2. For $\sigma_b=-\sigma_c$ we obtain

\ben
{}_sN_{m,\sigma_a,\sigma_b,-\sigma_b}+1= \pm\,m \sigma_b  -\sigma_a s \geq 1.
\la{sNm_Kerr_A}
\een

As seen, in the case of TAE the only function of the $\delta$-condition is to relate
the degree $N$ of the $\Delta_{N+1}$-condition with the spin-weight $s$ and the azimuthal number $m$.

\section{Polynomial Solutions of the TAE}
It can be shown that the polynomial solutions ${}_s S^{+}_{\omega,E,m,\sigma_a,\sigma_b,\sigma_c}$
are regular at the south pole and certainly singular at the north pole.
For the polynomial solutions ${}_s S^{-}_{\omega,E,m,\sigma_a,\sigma_b,\sigma_c}$ we have a similar
result, but with regularity at north pole and singularity at south one.

Studying the polynomial solutions to the TAE one must take into account one new obstacle.
The HeunC-factors in solutions \eqref{TAE16local_cc} do not become polynomials for negative
integer $\sigma_b b_{{}_\pm}< 0$ and for otherwise arbitrary values of the parameters,
because of the presence of $\ln(z)$ terms in equation \eqref{ccH}.
Hence, in contrast to the previous cases of RWE and TRE,
looking for polynomial solutions to the TAE we must impose the additional requirement
$\sigma_b b_{{}_\pm}\geq 0$ in the case of integer spin weights $s=0,\pm1,\pm2$.

Using the relations (\ref{munu:b}) and  (\ref{TAEparameters}) we obtain the general formula for the constant $E$ in the form
\ben
E^\pm\!=\!\mu^\pm\!-\!a\omega^2 \mp 2\sigma_a \big(1\!\mp\sigma_b m+(\sigma_a+\sigma_b)s\big) a\omega +
{\frac{\sigma_b\!-\sigma_c}{2}}m\left(\sigma_b m \mp 1\right) +
{\frac{\sigma_b\!+\sigma_c}{2}}s\left(\sigma_b s\!+1\right).
\la{E_TAE}
\een

Further analysis shows that we have again two classes of polynomial solutions to the TAE,
as in the cases of RWE and TRA, but their structure in some cases may be different.

\subsection{First Class of Polynomial Solutions to the TAE:}

These are the solutions ${}_sS_{\omega,E,m,-\sigma,-\sigma,-\sigma}^{\pm}$ with $\sigma=\text{sign}(s)$.
For them the condition (\ref{sN_Kerr_A}) is fulfilled independently of the values of the integer $m$, but
for integer $|s|$ the specific requirement $\sigma_b b_{{}_\pm}\geq 0$ yields the restriction $|m|\geq|s|$.
As in the previous cases of first class polynomial solutions (See Sections 3 and 6.) the value $s=0$ is eliminated
by (\ref{sN_Kerr_A}).
Hence, we have an infinite series of first class polynomial solutions to the TAE for all admissible values of $s$ and $m$.
Preserving the accepted in the previous sections style we denote the polynomial
solutions to TAE of the first class as ${}_sS_{\omega,E,m}^{\pm}={}_sS_{\omega,E,m,-\sigma,-\sigma,-\sigma}^{\pm}$.

For them the $\Delta_{N+1}$-condition reads $\Delta_{2|s|}(\mu)=0$ and has $2|s|$-in-number
solutions ${}_s\mu^\pm_{\omega,k,m}$. From the formulae (\ref{E_TAE}) one obtains
\ben
{}_s E^\pm_{\omega,k,m}={}_s\mu^\pm_{\omega,k,m} +|s|(|s|-1) -a\omega(a\omega-2m) \mp 2\sigma(2|s|-1)a\omega,
\la{E_first_classA}
\een
where $k=1,\dots,2|s|$, $s=\pm 1/2,\pm 1, \pm 3/2, \pm 2$ and for integer $|s|$ in addition $|m|\geq|s|$.

Solving the $\Delta_{N+1}$-condition, we obtain for the different values of $|s|$ as follows:
\ben
{}_sE_{\omega,m}^{\pm}=-a\omega(a\omega-2m)-{\frac 1 4}:\,\,\,\text{for}\,\,\,|s|={\frac 1 2}.
\la{E_first_class_1/2_A}
\een
\ben
{}_sE_{\omega,k,m}^{\pm}=-a\omega(a\omega-2m)-2(-1)^k\sqrt{a\omega(a\omega-m)}:\,\,\,\text{for}\,\,\,|s|=1,\,\,|m|\geq 1,\,\,k=1,2.
\la{E_first_class_1_A}
\een

Note that the values \eqref{E_first_class_1/2_A} and \eqref{E_first_class_1_A} of the separation constant $E$
obtained for the first class polynomial solutions to TAE are the same as
the corresponding values \eqref{E_first_class_1/2} and \eqref{E_first_class_1}
for the first class polynomial solutions to TRE.
The author's attention to this fact was drawn by Dr. Roumen Borissov
during the recent discussions on the applications of Heun's functions to TME.
The important consequences will be considered in a separate article \cite{RBPF}.

For the gravitational waves ($|s|=2$) the quantities
${}_s\mu_{\omega,k,m}^{\pm}$ are solutions of the algebraic equations of fourth degree $\Delta^\pm_4(\mu)=0$.
We do not need here the exact form of these roots. It is quite complicated.
Below we present only the form of the separation constant $E$ for TAE, obtained
making use of the Taylor series expansions of the roots around the point $a\omega=0$.

Thus we obtain for $|s|=2$ and $k=1,2$ the following eight values:
\ben
{}_sE_{\omega,k,m}^{\pm}=2-4ma\omega -i(-1)^k12\sqrt{(m-1)m(m+1)}\,(a\omega)^{3/2}
+6\left(m^2-{\frac 7 6}\right)(a\omega)^2
\!+\!{\cal{O}}_{5/2}(a\omega),\hskip .truecm\label{mu:1_2A}
\een
and for $|s|=2,\,\,m\neq 0,\,\,k=3,4$ another eight values:
\ben
{}_sE_{\omega,k,m}^{\pm}\!=\!-(-1)^k4\sqrt{m a\omega}\left(1+\left(3m-{\frac 2 m}\right)a\omega+{\cal{O}}_{2}(a\omega)\right)\!
+\!8ma\omega\!-\!6\!\left(\!m^2\!-\!{\frac 5 6}\!\right)\!(a\omega)^2\!+\!{\cal{O}}_{3}(a\omega).
\hskip .5truecm\label{mu:3_4A}
\een

As seen, for gravitational waves of first polynomial class the values \eqref{mu:1_2A} and \eqref{mu:3_4A}
of the corresponding constants $E$ differ substantially from the analogous values \eqref{mu:1_2} and \eqref{mu:3_4}
of the constants $E$, obtained for TRE in Section 6.1.1.
This is in sharp contrast to the case of neutrino waves ($|s|=1/2$) of first polynomial class
and to the case of electromagnetic waves ($|s|=1$) of this kind.

It can be shown that this phenomenon reflects the difference between the Starobinsky constants
for solutions with spin $|s|=2$ to TAE and TRE \cite{Teukolsky, Chandra}. For spins $|s|=1/2$ and $1$
the Starobinsky constants for solutions to TAE and TRE are the same.

Despite the pointed difference, the first-polynomial-class-solutions to TAE and TRE
with spin $|s|=2$ have similar qualitative properties, discussed at the end of Section 6.1.1.

\subsection{Second Class of Polynomial Solutions to TAE:}
We have a finite number of second class polynomial solutions to the TAE.
For them the conditions $N\geq 0$ and $\sigma_b b_{{}_\pm}\geq 0$ must be satisfied simultaneously.
Altogether there exist 24 such solutions ${}_sS_{\omega,E,m,\mp,\pm,\mp}^{\pm}$ :
\begin{eqnarray}
&{}_sS_{\omega,E,m,-,+,-}^{+}:&   s\!=\!+2,\,\, m\!=\!-1,\,\,\,\,0,\,1,\,2;\,\,  s\!=\!+1,\,\, m\!=\!\,\,\,\,0,\,1; \nonumber \\
&{}_sS_{\omega,E,m,+,-,+}^{+}:&   s\!=\!-2,\,\, m\!=\!-2,-1,\,0,\,1;\,\,         s\!=\!-1,\,\, m\!=\!-1, 0;  \nonumber\\
&{}_sS_{\omega,E,m,-,+,-}^{-}:&   s\!=\!+2,\,\, m\!=\!-2, -1,\,0,\,1;\,\,        s\!=\!+1,\,\, m\!=\!-1,\,0; \nonumber\\
&{}_sS_{\omega,E,m,+,-,+}^{-}:&   s\!=\!-2,\,\, m\!=\!-1,\,\,\,\,0,\,1,\,2;\,\,  s\!=\!-1,\,\, m\!=\!\,\,\,\,\,0,\,1.
\label{Second_Class_Polynomial_TAE}
\end{eqnarray}
The relation between the constants $E$ and $\omega$ follows from \eqref{E_TAE},
when $\mu$ in it is replaced by the solutions
of the $\Delta_{N+1}$-condition in the form $\Delta_{|s\pm m|}^\pm(\mu)=0$.

\section{The 256 Classes of Exact Solutions to the Teukolsky Master Equation}
Combining the studied in the previous Sections solutions to the TRE and TAE we can construct the following 256 classes of
exact solutions to the TME
\ben
{}_s\Psi^{\pm,\pm}_{\omega,E,m,\sigma_\alpha,\sigma_\beta,\sigma_\gamma,\sigma_a,\sigma_b,\sigma_c}(t,r,\theta,\varphi)=
e^{-i\omega t} e^{ im\varphi}
{}_sR^\pm_{\omega,E,m,\sigma_\alpha,\sigma_\beta,\sigma_\gamma}(r;r_{+},r_{-}) {}_sS_{\omega,E,m,\sigma_a,\sigma_b,\sigma_c}^{\pm}(\theta).
\la{256_Sol_TME}
\een

For specific physical problems one has to impose specific additional conditions,
like stability conditions, boundary conditions,
casuality conditions, specific fixing of the in-out properties, regularity conditions etc.
Thus one selects some specific combinations of solutions
to TRE and TAE in \eqref{256_Sol_TME} and derives
the spectrum of the separation constants $\omega$ and $E$ in the given problem.

For example, choosing solutions to TRE which enter both event horizon and 3D-space infinity we
are studying Kerr black holes. If in addition we choose regular solution to TAE, we will obtain
QNM of KBH. The choice of one-way polynomial solutions to the TAE in combination with BH
boundary conditions for TRE will produce jets from KBH \cite{PFDS, PFDS_astroph:HE}.
The combination of outgoing one-way  polynomial solutions of TRE and regular solutions of TAE
seem to be proper for description of supernovae explosions and may produce the structure
of their outbursts. The use of one-way polynomial solutions both for TRE and TAE \cite{RBPF}
seems to be most natural for description of jets from object, different from KBH.
Constructing simple models of different kinds of compact rotating relativistic objects
one can use Dirichlet's, or semi-Dirichlet's  boundary conditions on proper surface
outside the event horizon, or even outside the ergosphere of Kerr metric, i.e. acting
by analogy with the construction of such models in Schwarzschild space-time \cite{F}.
Excluding the ergoregion of Kerr metric from physical consideration we may ensure
the stability of the corresponding objects, which otherwise may be problematic
in some domain of parameters \cite{Visser}.
The exact polynomial solutions to the TRE of equidistant spectrum may be
useful for quantum gravity, etc.

The solutions \eqref{256_Sol_TME} do not necessarily have a direct physical meaning. Instead,
proper linear combination of specific solutions, which obey the corresponding boundary conditions,
is to describe the Nature. In general the solutions \eqref{256_Sol_TME} have to be considered
as an auxiliary mathematical objects -- (maybe singular) kernels of integral representations
\eqref{IntRepresentation} of the physical solutions.
The choice of the corresponding amplitudes ${}_sA_{\omega,E,m,\sigma_\alpha,\sigma_\beta,\sigma_\gamma,\sigma_a,\sigma_b,\sigma_c}$
will fix completely the physical model and can ensure the convergence of the integrals and discrete sums
to physically acceptable solutions.

\section{Conclusion}
We have demonstrated that the confluent Heun's functions are the adequate
and natural tool for unified description of the linear
perturbations to the gravitational field of Schwarzschild and Kerr metrics
outside the corresponding horizons, as well as in the interior domains. These functions
give us an effective tool for exact mathematical treatment of different boundary problems
and corresponding physical phenomena.

Large classes of exact solutions of all possible types to the perturbation
equations, both of Schwarzschild and Kerr metrics, are described and classified uniformly
in terms of confluent Heun functions
and confluent Heun polynomials. Using these functions we have re-derived the known polynomial
solutions and found  a large number of new ones.

We have to stress especially the newly obtained singular polynomial solutions of Teukolsky
angular equation. These solutions may describe in the most natural way the collimation of
the observed relativistic jets, related with different kind of astrophysical objects.

The solutions of the remaining basic mathematical problems and specific physical applications of the
obtained results will be published elsewhere.

\vskip .7truecm
\noindent{\bf \Large Acknowledgments}
\vskip .3truecm

I am thankful for the stimulating discussion  on exact solutions to
Regge-Wheeler and Teukolsky equations and different boundary problems
to Kostas Kokkotas, Luciano Rezzolla and Edward Malec
-- during the XXIV Spanish Relativity Meeting, E.R.E. 2006, to participants
in the seminar of the Astrophysical Group of the Uniwersytet Jagiellonski,
Institut, Fizyki, Cracow, Poland, May 2007,
to participants of the Conference Gravity, Astrophysics and Strings at Black Sea 2007,
Primorsko, Bulgaria and to the participants
in the seminar of the Department of Physics, University of in Nis, Serbia, December 2007.
The author is thankful too to Denitsa Staicova and Roumen~Borissov for numerous
discussions during the preparation of the present article,
and to Shahar Hod -- for his kind help in the enriching of the references.

I would like to express my gratitude to Professor Saul Teukolsky for his comments of the present article
and his useful suggestions.

This article was supported by the Foundation "Theoretical and
Computational Physics and Astrophysics" and by the Bulgarian National Scientific Found
under contracts DO-1-872, DO-1-895 and DO-02-136.

\end{document}